\documentclass[%
 aip,
 amsmath,amssymb,
 reprint,%
]{revtex4-1}

\usepackage{graphicx}% Include figure files
\usepackage{dcolumn}% Align table columns on decimal point
\usepackage{bm}% bold math
\usepackage[utf8]{inputenc}
\usepackage[T1]{fontenc}
\usepackage{mathptmx}
\usepackage{etoolbox}

\makeatletter
\def\@email#1#2{%
 \endgroup
 \patchcmd{\titleblock@produce}
  {\frontmatter@RRAPformat}
  {\frontmatter@RRAPformat{\produce@RRAP{*#1\href{mailto:#2}{#2}}}\frontmatter@RRAPformat}
  {}{}
}%
\makeatother
\begin{document}

\preprint{AIP/123-QED}

\title[Hamiltonian description for magnetic field lines: a tutorial]{Hamiltonian description for magnetic field lines:\\ a tutorial}
% Force line breaks with \\
\author{R. L. Viana}
\email{viana@fisica.ufpr.br}
\affiliation{Department of Physics, Federal University of Paraná, 80060-000 Curitiba, PR, Brazil}
\affiliation{Institute of Physics, University of São Paulo, 05508-900 São Paulo, SP, Brazil}%%
\author{M. Mugnaine}%
\affiliation{Institute of Physics, University of São Paulo, 05508-900 São Paulo, SP, Brazil}%
\author{I. L. Caldas} 
\affiliation{Institute of Physics, University of São Paulo, 05508-900 São Paulo, SP, Brazil}%

\date{\today}% It is always \today, today,

\begin{abstract}
Under certain circumstances, the equations for the magnetic field lines can be recast in a canonical form, after defining a suitable field line Hamiltonian. This analogy is extremely useful for dealing with a variety of problems involving magnetically confined plasmas, like in tokamaks and other toroidal devices, where there is usually one symmetric coordinate which plays the role of time in the canonical equations. In this tutorial paper we review the basics of the Hamiltonian description for magnetic field lines, emphasizing the role of a variational principle and gauge invariance. We present representative applications of the formalism, using cylindrical and magnetic flux coordinates in tokamak plasmas.
\end{abstract}

\maketitle

\section{Introduction}
	
Creating and confining hot plasmas are the foundation of fusion studies \cite{miyamoto,abdullaev}. As the increase of the temperature helps to create the plasma, magnetic fields are able to confine it in suitable containers.  The spatial structure of magnetic field lines is an important ingredient in many theoretical analyses of magnetically confined plasmas in toroidal devices like tokamaks, stellarators, reversed field pinches, etc. \cite{morosov}. In tokamaks, the magnetic field responsible for the confinement results from the superposition of the toroidal field, generated by external coils wound around the entire torus, and the poloidal field, due to plasma current itself \cite{wesson}. However, the equilibrium magnetic field can be modified by plasma oscillations or by external coils used to control instabilities \cite{evans}.
	
An interesting situation is where the magnetic field is time-independent, as the case in MHD equilibrium configurations \cite{miyamoto}. Starting from a symmetric plasma equilibrium configuration with an ignorable coordinate (e.g. the toroidal angle in tokamaks), the magnetic field line equations can be cast in the form of canonical equations, if the ignorable coordinate plays the role of time \cite{bernardin}. Furthermore, as the magnetic field is divergence free, we can describe the field lines using a two dimensional area-preserving map, with respect to a surface of section of the torus at a fixed toroidal angle \cite{morrison}. The resulting phase space of the field lines is identical to a Hamiltonian phase space, indicating that the field lines act, at least locally, as trajectories \cite{morrison}. Hence, the dynamics described by the corresponding Hamiltonian represents not a true motion but instead a magnetostatic structure parameterized by the time-like coordinate \cite{cary}. The main advantage of making this analogy is to use the powerful toolbox of Hamiltonian theory to investigate the magnetic field line structure, specially if non-symmetrical perturbations are considered \cite{goldstein}.  
	
The analogy between magnetic field lines and a Hamiltonian system has been first pointed out by Kruskal, in 1952 \cite{morrison,escande}. Kruskal proposed and iterated an area-preserving map, similar to the standard map, in order to describe the magnetic field lines of stellarators \cite{morrison,escande}. This connection between field lines and Hamiltonian formalism was also recognized simultaneously but independently in United States by Donald Kerst (the inventor of Betatron)  \cite{kerst} and in Soviet Union \cite{gelfand}. Nevertheless, an explicit and generalized Hamiltonian description was only proposed later, by Whiteman \cite{whiteman} and Boozer \cite{morrison,escande,evans}.
	
Even though magnetic field lines can be described by low-dimensional Hamiltonian systems, the numerical integration of the motion equations can be computationally costly \cite{escande2}. For this reason, explicit area-preserving maps, derived from the Poincaré map of the magnetic field line, are often used. These Hamiltonian maps are an important tool for studying the kinetic and fluid transport process as plasma turbulence and MHD stability, once they inform us about the global and fine scale structure of the edge magnetic topology in toroidal systems \cite{evans}. Besides, these maps permit long time examination of individual trajectories for the statistical analysis of the field lines and the investigation of transport with a reasonable computational time \cite{abdullaev}. 
	
The theory of magnetic field lines in confined plasma devices could not guarantee the regularity of the field lines \cite{escande}. An example is the study of magnetostatic perturbations produced by coils placed outside the plasma: the resulting magnetic field lines may present unexpected and complicated behaviors like periodic, quasi-periodic, and even chaotic orbits \cite{morrison,firpo}. The latter, in particular, represent a local destruction of the magnetic surfaces that confine the plasma \cite{filonenko,filonenko1}.

The main motion of the plasma particles is along to the field lines while slowly drifting  across the equilibrium fields due to the lines curvature, the particle rotation around the lines and electric drift\cite{hazeltine,horton,cary2009}. Thus, as the fastest motion is along the field lines, the particle escape to the wall can be predicted by the field line configuration \cite{abdullaev2}. Observing magnetic field lines in a surface section of a tokamak, they can be closed lines within the trace of a toroidal magnetic surface or they can fill a two dimensional domain. For the first case, the field line is regular, while the second case indicate chaos \cite{escande}. Chaotic behavior is related to the topology of the magnetic field lines and the dynamics of the particles gyrating along these lines, as well as the turbulent transport, ray dynamics and radiofrequency heating \cite{escande,escande2}. Furthermore, the non uniform particle transport at the tokamak plasma edge can be roughly estimated from the field lines escaping to the wall    \cite{abdullaev2,jakubowski}. Broader reviews between Hamiltonian chaos and fusion plasmas can be found in Refs. \cite{escande,escande2,evans}.
	
From a classic mechanics framework, a general description in curvilinear coordinate system was proposed by Whiteman \cite{whiteman}, and  extended later by Boozer \cite{boozer}, Cary and Littlejohn \cite{cary}, and Elsässer \cite{elsasser}. The general formalism described by Whiteman has been applied to a variety of coordinate systems: cylindrical \cite{viana1}, helical \cite{viana2}, spherical \cite{viana3}, and pseudo-toroidal \cite{viana4}. Various applications of the magnetic field line Hamiltonian have been made by Freis {\it et al.} \cite{licht2} and Hamzeh \cite{hamzeh} for a toroidal machine called {\it Levitron}, and by Lichtenberg \cite{licht,licht3} in an investigation of the $m = 1$ island on sawtooth oscillations in Tokamaks. 

Besides its applications in fusion plasmas, the Hamiltonian description of magnetic field lines provides a nice non-mechanical example of the usefulness of the Hamiltonian formalism to intermediate and advanced students. Moreover, the magnetic field line problem has the unique feature that the corresponding phase space actually coincides with the configuration space, what facilitates the visualization of complex dynamical concepts like KAM tori, homoclinic tangles, and so on. On the other hand, the basic material on the Hamiltonian description of magnetic field lines is often available only in publications targeted to the plasma physics experts, what creates an additional difficulty for an interested novice reader.
	
In order to overcome the latter problem we wrote this tutorial as an aid to students and researchers interested to master the basic ideas of the Hamiltonian description for the magnetic field lines. Moreover, we present some representative applications of this formulation so as to illustrate its usefulness in plasma physics problems. We emphasize that this paper is not a review of this subject. While we focused on fusion plasmas, the methods can also be used in plasmas of astrophysical and geophysical interest, provided we have situations of MHD equilibrium with adequate stability properties. 

This paper is organized as follows: in Section II we show the derivation of a variational principle for magnetic field lines and the role played by gauge invariance. The Hamiltonian description in general curvilinear coordinates is presented in Section III. In Section IV we describe in some detail an application of the description in cylindrical coordinates to a large aspect-ratio tokamak with an ergodic magnetic limiter, using canonical perturbation theory to derive an analytical formula for the width of magnetic islands, which is an expression of practical interest for stability and transport theoretical studies of tokamak plasmas \cite{miyamoto}. In Section V we present an application of the general formulation for magnetic flux coordinates, which are widely in numerical codes for computer simulation of plasmas \cite{boozer}, displaying an application to a magnetic field line map (tokamap) proposed by Balescu and coworkers \cite{balescu}. The last section is devoted to our Conclusions. 
	
\section{Variational principle}
	
The equations of motion of a particle can be derived from Hamilton´s variational principle \cite{goldstein}
	\begin{equation}
		\label{hamprinc}
		\delta \int_{t_1}^{t_2} dt \, L(q,{\dot q},t) = 0,
	\end{equation}
where $L$ is the Lagrangian, $q$ and ${\dot q}$ are the generalized coordinate and velocity, respectively, and $t_{1,2}$ are fixed instants of time. It means that, considering the infinite possible paths connecting the particle positions at fixed times $t_{1,2}$, the actual trajectory between them is that for which the integral $\int L dt$ is an extremum. The exploitation of this principle gives the Euler-Lagrange equations of motion for the particle. 
	
For a non-relativistic particle with mass $m$ and charge $e$, subjected to electromagnetic fields, the Lagrangian is 
	\begin{equation}
		\label{lagfields}
		L = \frac{1}{2}  \, m v^2 - e  \, \Phi + e  \, {\bf A}\cdot{\bf v},
	\end{equation}
where $\Phi$ and ${\bf A}$ are, respectively, the scalar and vector potentials, from which the electric and magnetic fields are given by \cite{jackson}
	\begin{equation}
		\label{EBfields}
		{\bf E} = - \nabla\Phi - \frac{\partial{\bf A}}{\partial t}, \qquad {\bf B} = \nabla\times{\bf A}.
	\end{equation}
	
The scalar and vector potentials do not determine uniquely the electromagnetic fields: the latter are invariant under gauge transformations
	\begin{equation}
		\label{gauget}
		\Phi \rightarrow \Phi + \frac{\partial\chi}{\partial t}, \qquad {\bf A} \rightarrow {\bf A} - \nabla\chi,
	\end{equation}
where $\chi({\bf r},t)$ is an arbitrary function. 
	
The variational principle for the magnetic field lines can be obtained from Hamilton´s principle (\ref{hamprinc}) by considering a massless particle under a pure magnetic field, i.e. $\Phi = 0$:
	\begin{equation}
		\label{hamfield1}
		\delta \int_{t_1}^{t_2} {\bf A}\cdot{\bf v}  \, dt = 0,
	\end{equation}
	or, changing the integration from time to space
	\begin{equation}
		\label{hamfield}
		\delta \int_{{\bf r}_1}^{{\bf r}_2} {\bf A}\cdot{\bf dr} = 0,
	\end{equation}
where ${\bf r}_{1,2}$ are fixed spatial positions \cite{morosov,cary}. The variational principle (\ref{hamfield}) states that, considering the infinite paths in space connecting the fixed points ${\bf r}_1$ and ${\bf r}_2$, the magnetic field line is the path for which the integral $\int {\bf A}\cdot{\bf dr}$ is an extremum, for a given time. Recently, Escande and Mono have introduced a novel approach to the variational principle (\ref{hamfield}) using Stokes theorem \cite{momo}. This procedure allows a general treatment of some problems, like determining the width of magnetic islands, a subject that will be address in section IV.
	
In the following, we will use the Einstein sum convention for repeated indices and express a vector using their contravariant and covariant components, as well as the corresponding basis vectors, in the forms,
	\begin{equation}
		\label{contraco}
		{\bf A} = A_i  \, {\hat{\bf e}}^i, \qquad 
		{\bf dr} = dx^j  \, {\hat{\bf e}}_j.
	\end{equation}
Since contravariant and covariant basis vectors are dual, \textit{i.e.},
	\[
	{\hat{\bf e}}^i \cdot {\hat{\bf e}}_j = \delta^i_j,
	\]
we can rewrite the variational principle for field lines (\ref{hamfield}) in the form 
	\begin{equation}
		\label{hamfield2}
		\delta \int_{1}^{2} A_i \, dx^i = 0.
	\end{equation}
	
It is useful to introduce here a variational parameter $\lambda$ which, as in mechanics, labels the infinite possible paths connecting the fixed points in Hamilton´s principle. Each path is thus represented by a function of this variational parameter 
	\begin{equation}
		\label{xis}
		x^1 = x^1(\lambda), \qquad x^2 = x^2(\lambda), \qquad x^3 = x^3(\lambda).
	\end{equation}
	
The variational principle (\ref{hamfield2}) can be written as an integral over $\lambda$
	\begin{equation}
		\label{varlambda}
		\delta \int_{\lambda_1}^{\lambda_2} d\lambda \left( A_1  \, \frac{dx^1}{d\lambda} + A_2  \, \frac{dx^2}{d\lambda} + A_3  \, \frac{dx^3}{d\lambda} \right) = 0,
	\end{equation}
such that the functions $A_i(x^1,x^2,x^3)$ are fixed, but the arguments (\ref{xis}) have to be varied independently from $\lambda$, with vanishing variation $\delta x^i(\lambda)$ at the fixed points $\lambda_1$ and $\lambda_2$. As a consequence, the exploitation of this variational principle gives the equations for the magnetic field lines \cite{elsasser}
	\begin{equation}
		\label{fieldlines}
		\frac{dx^1}{B^1} = \frac{dx^2}{B^2} = \frac{dx^3}{B^3},
	\end{equation}
which are equivalent to the vector equation
	\begin{equation} 
		\label{vecfield}
		{\bf B} \times {\bf dr} = {\bf 0}.
	\end{equation}
	
We can choose a gauge such that one of the covariant components of the vector potential vanishes, e.g.
	\begin{equation}
		\label{a20}
		A_2 = 0,
	\end{equation}
and the variational principle reduces to 
	\begin{equation}
		\label{hamfield3}
		\delta \int_1^2 \left( A_1  \, dx^1 + A_3  \, dx^3 \right) = 0.
	\end{equation}
	
The variational parameter $\lambda$ is arbitrary, but it is convenient to choose it such that $\lambda$ is an ignorable coordinate, i.e. physically relevant quantities do not depend on it. In this case, magnetic field lines stream along the direction of this coordinate. A common choice in toroidal fusion devices like tokamaks is the azimuthal angle $x^3$. We thus take $\lambda = x^3$ and the variational principle (\ref{hamfield3}) becomes 
	\begin{equation}
		\label{varlambda1}
		\delta \int_{x^3_1}^{x^3_2} dx^3 \left( A_1  \, \frac{dx^1}{dx^3} + A_3 \right) = 0,
	\end{equation}
where $x^3_{1,2}$ are the values of the azimuthal angle at the fixed points. 
	
\section{Hamiltonian description}
	
Let us consider a dynamical system with one degree of freedom, whose state is described by a generalized coordinate $q$ and a generalized velocity ${\dot q}$, with Lagrangian $L(q,{\dot q},t)$. The generalized momentum $p$ canonically conjugated to the coordinate $q$ is given by $p = \partial L/\partial{\dot q}$. In terms of the latter, the modified Hamilton's principle is written as 
	\begin{equation}
		\label{varham1}
		\delta\int_{t_1}^{t_2} dt \, \left\{ p  \, {\dot q} - H(p,q,t) \right\} = 0,
	\end{equation}
where $H$ is the system Hamiltonian. We rewrite this expression as \begin{equation}
		\label{varham}
		\delta \int_{1}^{2} \left\{ p \, dq - H(p,q,t) \, dt \right\} = 0,
	\end{equation}
where $1$ and $2$ represent fixed points, as before. 
	
Comparing (\ref{varham}) with the variational principle (\ref{hamfield3}) for magnetic field lines, we can make the following associations 
	\begin{eqnarray}
		\label{q}
		q & = & x^1, \\
		\label{p}
		p & = & A_1(x^1,x^2,x^3), \\
		\label{t}
		t & = & x^3, \\
		\label{H}
		H & = & - A_3(x^1,x^2,x^3).
	\end{eqnarray}
Hence, magnetic field lines can be described as a Hamiltonian system, where the role of time is played by the ignorable  coordinate $x^3$, since the "physical" time is kept strictly fixed. 
	
In this description, the magnetic field line equations (\ref{fieldlines}) can be written as Hamilton´s equations
	\begin{align}
		\label{hameq1}
		\frac{dq}{dt} & = \frac{\partial H}{\partial p}, \\
        \label{hameq2}
		\frac{dp}{dt} & = - \frac{\partial H}{\partial q},
	\end{align}
for an one-degree-of-freedom system described by the canonical pair of variables $(p,q)$. If the field line Hamiltonian $H$ does not depend explicitly on time $t = x^3$, as in axisymmetric plasma equilibrium configurations, the Hamiltonian $H(p,q)$ describes an integrable system, being a constant of "motion", or a first integral. On the other hand, magnetic perturbations caused by external fields or instabilities can introduce non-axisymmetric terms in the Hamiltonian, resulting on a time-dependent system $H(p,q,t)$, which is generally non-integrable. 
	
Instead of the vector potential, we can obtain a Hamiltonian description directly from the magnetic field components. Writing the relation ${\bf B} = \nabla\times{\bf A}$ in curvilinear coordinates we have
	\begin{eqnarray}
		\label{B1}
		B^1 & = & \frac{1}{\sqrt{g}} \left( \frac{\partial A^3}{\partial x^2} - \frac{\partial A^2}{\partial x^3} \right), \\
		\label{B2}
		B^2 & = & \frac{1}{\sqrt{g}} \left( \frac{\partial A^1}{\partial x^3} - \frac{\partial A^3}{\partial x^1} \right), \\
		\label{B3}
		B^3 & = & \frac{1}{\sqrt{g}} \left( \frac{\partial A^2}{\partial x^1} - \frac{\partial A^1}{\partial x^2} \right),
	\end{eqnarray}
where $g = \det g_{ij}$ is the determinant of the covariant metric tensor, whose elements are given by $g_{ij} = {\hat{\bf e}}_i \cdot {\hat{\bf e}}_j$. If the non-diagonal elements of the metric tensor are identically zero, the coordinate system is called orthogonal. In this case the diagonal elements are also called metric coefficients, and we have that $g = g_{11} g_{22} g_{33}$.
	
After choosing a gauge for which $A_2 = 0$ these expressions lead  to the remaining components of the vector potential
	\begin{eqnarray}
		\label{A1}
		A_1 & = & - \int \sqrt{g} B^3 dx^2, \\
		\label{A3}
		A_3 & = & \int \sqrt{g} B^1 dx^2. 
	\end{eqnarray}
In terms of the magnetic field components, the Hamiltonian description of field lines is \cite{whiteman,bernardin}
	\begin{eqnarray}
		\label{q1}
		q & = & x^1, \\
		\label{p1}
		p & = & -\int \sqrt{g} B^3 dx^2, \\
		\label{t1}
		t & = & x^3, \\
		\label{H1}
		H & = & -\int \sqrt{g} B^1 dx^2.
	\end{eqnarray}
According to Eq. (\ref{B2}), these definitions are subjected to the following relation
	\begin{equation}
		\label{relat}
		\sqrt{g} B^2 = \frac{\partial H}{\partial q} + \frac{\partial p}{\partial t},
	\end{equation}
which can also be regarded as a direct consequence of the magnetic Gauss' law $\nabla\cdot{\bf B} = 0$. Janaki and Ghosh have shown that, under suitable canonical transformations, other choices of canonical pairs can be used, corresponding to different gauge coordinates \cite{janaki}.
	
When using the above formulae, one must have in mind that quite often the contravariant components of the magnetic field have not the same dimensions as the field itself due to the metric coefficients. In order to avoid inconsistencies, it is better to use the "physical" components defined by $B_{<i>} = \sqrt{g_{ii}} B^i$, where no sum in the index $i$ is intended. In the forthcoming section, our aim is to present representative applications of this description, using different coordinates systems, as the cylindrical and magnetic flux coordinates.
	
\section{Cylindrical coordinates}
	
Let us consider a toroidal plasma with major radius $R_0$ and minor radius $a$. In the local (or pseudo-toroidal) system of coordinates $(r,\theta,\phi)$, $\theta$ and $\phi$ are the poloidal and toroidal angles, respectively, and $r$ is the radial distance to the magnetic axis, which is a circle of radius $R_0$ centered at the torus major axis. 
	
The torus aspect ratio is $\epsilon = R_0/a$. In the large aspect ratio approximation ($R_0 \gg a$) we can neglect the toroidal curvature and consider the torus as a periodic cylinder of radius $a$ and length $2\pi R_0$. In this case it is possible to use cylindrical coordinates $(r,\theta,z)$, where $z = R_0 \phi$ is the rectified toroidal circumference. Due to the periodicity, we identify all points for which $z$ is an integer multiple of $2\pi R_0$.
	
Identifying $x^1 = \theta$, $x^2 = r$, $x^3 = z$, we have the metric coefficients $g_{11} = r^2$, $g_{22} = g_{33} = 1$. The "physical" components of the magnetic field are $B_{<1>} = B_\theta = r B^1$, $B_{<2>} = B_r = B^2$, $B_{<3>} = B_z = B^3$. The Hamiltonian variables read \cite{viana1}
	\begin{eqnarray}
		\label{q1c}
		q & = & \theta, \\
		\label{p1c}
		p & = & -\int dr~ r B_z, \\
		\label{t1c}
		t & = & z = R_0 \phi, \\
		\label{H1c}
		H & = & -\int dr B_\theta.
	\end{eqnarray}
	
In the large aspect ratio approximation we usually suppose the following equilibrium magnetic field components \cite{wesson},
	\begin{equation}
		\label{mfc}
		B_r = 0, \qquad B_\theta = B_\theta(r), \qquad B_z = B_0 = const.
	\end{equation}
such that the magnetic surfaces $r = const.$ are coaxial cylinders. On each cylinder the magnetic field lines spiral such that, after a complete toroidal turn (which corresponds to a whole excursion of $2\pi R_0$ along the periodic cylinder) the corresponding value of its poloidal angle increases by an angle $\iota$, called rotational transform (in Stellarator literature). Hence
	\begin{equation}
		\label{rotrans}
		\frac{d\phi}{d\theta} = \frac{1}{R_0} \frac{dz}{d\theta} = \frac{2\pi}{\iota(r)},
	\end{equation}
where, in general, the rotational transform is different for each magnetic surface. 
	
In Tokamak literature, we use the so-called safety factor $q = 2\pi/\iota$. From the magnetic field line equations
	\begin{equation}
		\label{mfle}
		\frac{r d\theta}{B_\theta} = \frac{R_0 ~d\phi}{B_z},
	\end{equation}
such that, in the large aspect ratio approximation, 
	\begin{equation}
		\label{safety}
		q(r) = \frac{d\phi}{d\theta} = \frac{r B_0}{R_0 B_\theta(r)}. 
	\end{equation}
If the safety factor varies monotonically with $r$, we have a twist system. The derivative $dq/dr$ is also called magnetic shear. If there are radial positions for which $q(r)$ has extrema, the magnetic shear is zero at those positions, and the system will be non-twist, since the variation of $q(r)$ is non-monotonic \cite{morrison}. 
	
The canonical momentum (\ref{p1c}) is,
\begin{equation}
	\label{p1c1}
	p = -B_0 \int r dr = -\frac{1}{2} B_0 r^2,
\end{equation}
up to an unessential constant, whereas the Hamiltonian (\ref{H1c}) \begin{equation}
	\label{H1c1}
	H = -\frac{B_0}{R_0} \int \frac{r dr}{q(p)} = \frac{1}{R_0} \int \frac{dp}{q(p)}
\end{equation}
	
Finally, we perform a non-canonical transformation in order to rescale variables
\[
(p,q,t) \rightarrow \left( J = -\frac{p}{B_0} = \frac{r^2}{2}, \theta, \phi = \frac{t}{R_0} \right),
\]
and the new Hamiltonian is $H_0 = -R_0 H/B_0$, written as 
\begin{equation}
	\label{ham0}
	H_0 = \int \frac{dJ}{q(J)}.
\end{equation}
	
For an example, let us consider a plasma column of radius $a$, for which the electric current density ${\bf j}$ has radial symmetry with respect to the axis, and carrying a total plasma current $I_p$
\begin{equation}
	\label{8perfil}
	j_z(r) = j_0 \left( 1 - \frac{r^2}{a^2} \right),
\end{equation}
where $j_0 = 2 I_p/\pi a^2$. Applying Ampère´s circuital law we obtain the poloidal field radial profile 
\begin{equation}
	\label{8ampere1}
	B_\theta(r) = B_{\theta a} \frac{r}{a} \left( 2 - \frac{r^2}{a^2} \right),
\end{equation}
where $B_{\theta a} = \mu_0 I_P/(2\pi a)$. The corresponding safety factor (\ref{safety}) is given by 
\begin{equation}
	\label{8bp2}
	q(r) = q_a {\left( 2 - \frac{r^2}{a^2} \right)}^{-1},
\end{equation}
where $q_a = (2\pi a^2 B_0)/(\mu_0 R_0 I_P)$ such that, at the symmetry axis we have $q_0 = q_a/2$. 
	
Substituting (\ref{8bp2}) into (\ref{ham0}) gives the field line Hamiltonian 
\begin{equation}
	\label{8H0}
	H_0(J) = \frac{2J}{q_a} \left(1 - \frac{J}{2a^2}\right).
\end{equation}
The integrable equilibrium configuration is described by a ``time"- independent Hamiltonian $H_0(J)$, which canonical equations are:
\begin{eqnarray}
\begin{aligned}
    \dfrac{d\theta}{d\phi}&=\dfrac{\partial H_0}{\partial J},\\
    \dfrac{dJ}{d\phi}&=0.
\end{aligned} 
\end{eqnarray}
Hence $J$ is a constant and, from (\ref{safety}), the field lines are helices wound around invariant tori of radii $J$ according to their safety factors $q(J)$. If $q(J)=m/n$, where $m$ and $n$ are integers, the respective torus is called ``rational", otherwise the torus is irrational.

\subsection{Ergodic magnetic limiter}
	
Divertors are devices used in tokamak experiments with the purpose to displace the interactions between the plasma particles and the tokamak wall, thereby avoiding direct contact between them and improving plasma confinement \cite{wesson,elton}. Initially, the divertors were designed to act directly over the toroidal or the poloidal field of the plasma and they required additional coil currents of the magnitude of plasma currents or even larger \cite{karger}. This led to experimental limitations and technological problems for the tokamak operation \cite{karger}.  As an alternative, Karger and Lackner proposed the helical divertor, which requires  smaller currents and possesses helical symmetry, which generates a magnetic field that resonates with the field at a surface in the plasma boundary, which is diverted \cite{karger}. 
 
 The resonance created by the divertor can also lead to a chaotic motion in the plasma edge, a process called "ergodization" \cite{engelhardt,vannucci}. The term "ergodic", however, has been later replaced by "chaotic", which is a more adequate description of the area-filling orbit created when the invariant manifolds stemming from unstable periodic orbits intercept in a complicated way forming the homoclinic tangle \cite{sanjuan}. The chaotic field lines increase the diffusion coefficient at the boundary of the plasma, reducing the plasma contamination \cite{engelhardt}, controlling the MHD oscillations \cite{vannucci}, reducing thermal flux density \cite{mccool}, and  controlling plasma-wall interaction \cite{portela2003}. For more information about the theoretical and experimental development of helical divertors, confer the References \cite{karger,engelhardt,mccool,pulsator,grosman}.
	
	\begin{figure}
		\begin{center}
			\includegraphics[width=0.4\textwidth,clip]{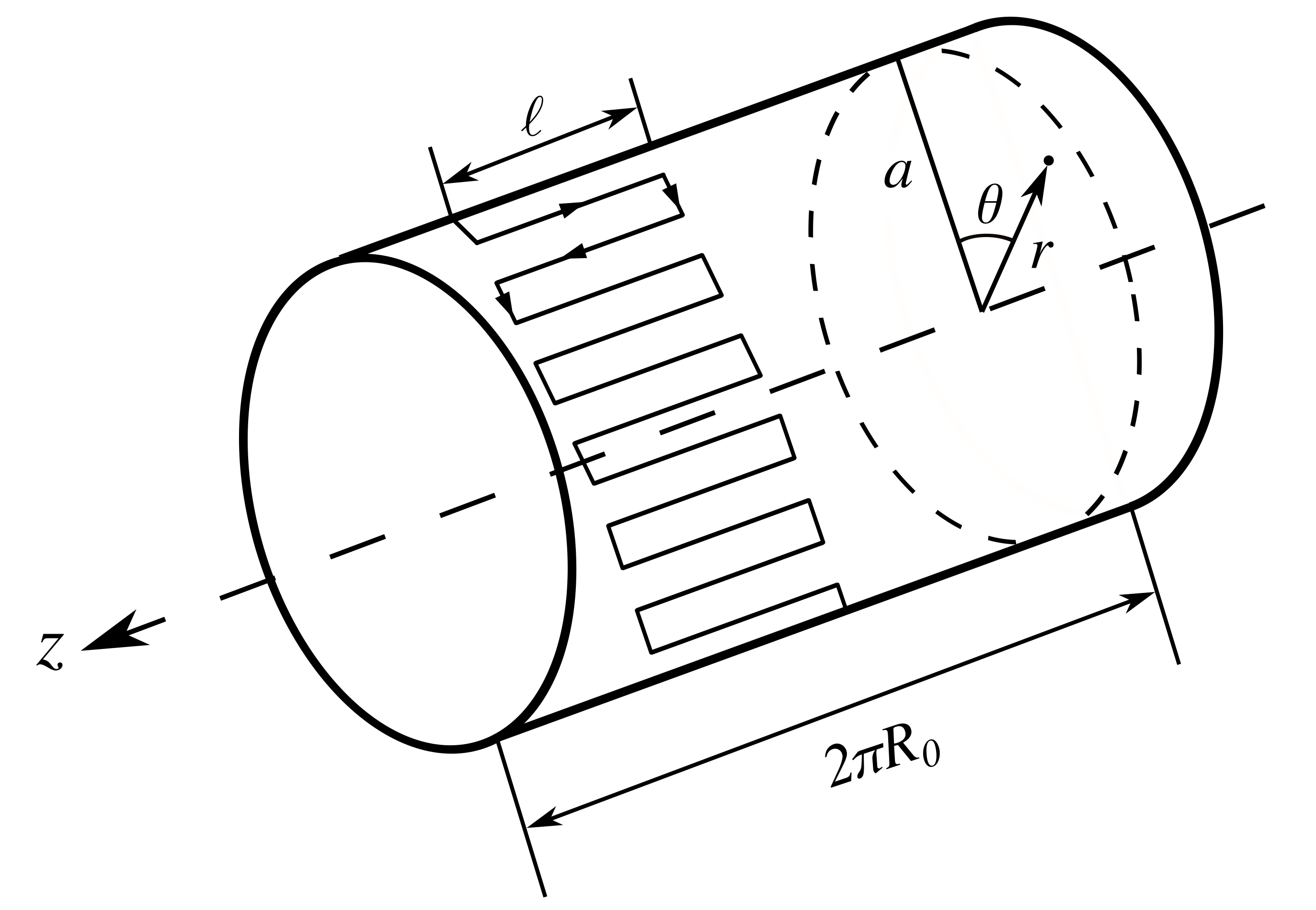}
			\caption{\label{limitador}  Schematic view of a ergodic magnetic limiter in a Tokamak with a large aspect ratio.}
		\end{center}
	\end{figure}
	
The external magnetic fields generated by the helical divertor create magnetic islands that can overlap and, consequently, form a stochastic layer in the plasma edge. Therefore, such a divertor is also called Ergodic Magnetic Limiter (EML) \cite{mccool}. The EML is a filamentary current ring with length $\ell$, wound around the torus (with radius $a$). There are two types of current segments in a EML (FIG. \ref{limitador}): straight segments parallel to the magnetic axis and curved segments along the poloidal direction. There are $m$ pairs of segments, such that two adjacent segments carry a current $I_L$ in opposite directions \cite{martin}, which produces a resonant helical field \cite{vannucci}.
	
Neglecting border effects, the magnetic field produced by an EML has the following components \cite{viana91}
	\begin{eqnarray} 
		\label{8Br1}
		B_r^{(1)}(r,\theta,\phi) & = & - \frac{\mu_0 m I_L}{\pi a^m} \, r^{m-1}  \, \sin(m\theta)  \, f(\phi), \\
		\label{8Bt1}
		B_\theta^{(1)}(r,\theta,\phi) & = & - \frac{\mu_0 m I_L}{\pi a^m}  \, r^{m-1}  \, \cos(m\theta)  \, f(\phi), 
	\end{eqnarray}
where the "time"-dependent factor is
	\begin{equation} 
		\label{8fator}
		f(\phi) = \begin{cases}
			1, & \text{if } 0 \le \phi < \ell/R_0, \\
			0, & \text{if } \ell/R_0 \le \phi < 2\pi,
		\end{cases} 
	\end{equation}          
whose Fourier decomposition, due to the $2\pi$-periodicity in the "time" $\phi$, is 
	\begin{equation}
		\label{8fator1}
		f(\phi) = \frac{\ell}{2\pi R_0} \left\{ 1 + 2  \, \sum_{n=1}^\infty \cos(n\phi) \right\}.
	\end{equation}
	
Changing to the action variable $J = r^2/2$ and using (\ref{8fator1}), the EML Hamiltonian reads
\begin{eqnarray}
\begin{aligned}
\label{8H1}
	H_1(J,\theta,\phi) =  - &\frac{\mu_0 R_0 I_L}{B_0 \pi a^m}  \, {(2J)}^{m/2}  \, \cos(m\theta)  \, f(\phi),  \\
	=  - &\sigma  \, A_m(J) \bigg\{ \cos(m\theta) + \\
    &  \, \left.\sum_{n=1}^\infty 
	\left\lbrack \cos(m\theta - n\phi) + \cos(m\theta + n\phi) \right\rbrack \right\},
\end{aligned}
\end{eqnarray}
where we define 
	\begin{eqnarray} 
		\label{8defsigma}
		\sigma & = & \frac{\mu_0 I_L \ell}{2 \pi^2 B_0} , \\
		\label{8defam}
		A_m(J) & = & \frac{{(2J)}^{m/2}}{a^m}.
	\end{eqnarray} 
	
The effect of an EML upon the equilibrium magnetic surfaces can be regarded as a quasi-integrable Hamiltonian system,
	\begin{equation}
		\label{combined}
		H(J,\theta,\phi) = H_0(J) + H_1(J,\theta,\phi),
	\end{equation}
if the perturbation strength is such that $|H_1| \ll |H_0|$. On defining the non-dimensional quantities
	\begin{equation}
		\label{8normas}
		\varepsilon = \frac{I_L}{I_P}, \qquad \xi = \frac{\ell}{R_0},
	\end{equation}
the perturbation strength (\ref{8defsigma}), given by,
	\begin{equation}
		\label{8sigma}
		\sigma = \varepsilon \xi \left( \frac{a^2}{q_a \pi} \right)
	\end{equation}
is small, for typical values of $a$ and $q_a$, provided $\varepsilon \ll 1$ and $\xi \ll 1$. 
	
In order to verify whether or not these conditions are satisfied we will take parameters from the TCABR Tokamak (Instituto de Física, Universidade de São Paulo, Brazil) presented in Ref. \cite{galvao}, which $R_0 = 0.61$m and $a = 0.18$ m. The plasma current is about $I_p = 0,1 MA$ and the toroidal field at the magnetic axis is $B_0 = 1,1 T$. The safefy factor is $q_a = 2,95 \approx 3$ at the plasma edge, and $q_0 = 1,5$ at the magnetic axis. An EML has been installed in TCABR with $m = 3$ pairs of wires with length $\ell = 0,1~m$, carrying a current about $I_L = 2500~A$ \cite{pires}. These values imply that $\varepsilon = 0,025$ and $\xi = 0,163$ are small enough to justify treating the EML field as a Hamiltonian perturbation upon the equilibrium magnetic surfaces. Ergodic limiters have also been used in other tokamaks as in Textor \cite{abdullaev}, Tore-Supra \cite{grosman}, and Text \cite{mccool}.
	
\subsection{Resonances and the pendulum approximation}
	
A resonance occurs wherever the phase $m\theta - n \phi$ is constant with respect to the "time" $\phi$, such that 
\begin{equation} 
	\label{8resonance}
	%	m \frac{d\theta}{d\phi} - n = 0 \rightarrow 
  \dfrac{d\phi}{d\theta} = \dfrac{m}{n}.
\end{equation}
From (\ref{safety}) it follows that, at the radial location $r^*$ of a given resonance, the safety factor is a rational number. The respective action variable is 
\begin{equation}
	\label{8acao}
	J^* = a^2 \left(1 - \frac{n q_a}{2m} \right),
\end{equation}
and it is also the position of a rational torus.

Near the exact resonance position the term $\cos(m\theta - n\phi)$ slowly oscillates, whereas all the remaining terms in the Fourier expansion vary rapidly with time and vanish if an average is performed over $\phi$. There remains only the resonant term, which reads 
	\begin{equation}
		\label{8hamtotal1}
		H_{res}(J,\theta,\phi) = H_0(J) - \sigma A_m(J) \cos(m\theta - n\phi).
	\end{equation}
	
According to Poincaré-Birkhoff theorem, all resonant (rational) tori are destroyed under a non-integrable perturbation, leaving in their places an even number of fixed points, half of them elliptic and half hyperbolic ones \cite{lichtenberg}. Let us concentrate our attention on some elliptic point at a given rational tori with $q = m/n$. Expanding the resonant Hamiltonian in the vicinity of $J = J^*$ we have, in powers of the small difference $\Delta J = J - J^*$, that 
\begin{widetext}
	\begin{eqnarray} 
		\nonumber
		H_{res}(J,\theta,\phi) & = & H_0(J^* + \Delta J) - \sigma A_m(J^* + \Delta J) \cos(m\theta - n\phi), \\
		\label{8hamtotal2}
		& = & H_0(J^*) + \Delta J {\left(\frac{\partial H_0}{\partial J}\right)}_{J^*} + \frac{{(\Delta J)}^2}{2} {\left(\frac{\partial^2 H_0}{\partial J^2}\right)}_{J^*} - \sigma A_m(J^*) \cos(m\theta - n\phi),
	\end{eqnarray} 
 \end{widetext}
in such a way that, using (\ref{8H0}), the Hamiltonian describing the motion near a resonance is 
	\begin{equation}
		\label{8hamdef}
		\begin{aligned}
			\Delta H(\Delta J, \theta,\phi) &= H_{res}(J,\theta,\phi) - H_0(J^*) \\
			&=\frac{n}{m}  \, \Delta J - \frac{1}{q_a a^2}  \, {(\Delta J)}^2-\sigma A_m(J^*) \cos(m\theta-n\phi).
		\end{aligned}
	\end{equation}
	
Performing a canonical transformation $(\Delta J,\theta,\phi) \rightarrow (I,\psi)$ using the generating function 
	\begin{equation}
		\label{8f2}
		F_2(I,\theta,\phi) = (m\theta - n\phi) I,
	\end{equation}
a straightforward calculation gives the pendulum Hamiltonian 
	\begin{equation}
		\label{8pendulo}
		{\cal H}(I,\psi) = \frac{1}{2}  \, G I^2 - F \cos\psi,
	\end{equation}
where $\psi=m\theta - n\phi$ and
	\begin{eqnarray} 
		\label{8F}
		F & = & \sigma  \,  A_m(J^*) = \frac{\sigma}{a^m}  \, {(2J^*)}^{m/2} = \sigma {\left\{ 2 \left(1 - \frac{n q_a}{2 m} \right) \right\}}^{m/2}, \\ 
		\label{8G}
		G & = & m^2 {\left(\frac{\partial\pounds}{\partial J}\right)}_{J^*} = - \frac{2m^2}{q_a a^2}.
	\end{eqnarray} 
	
\begin{figure}[!h]
	\begin{center}
		\includegraphics[width=0.5\textwidth]{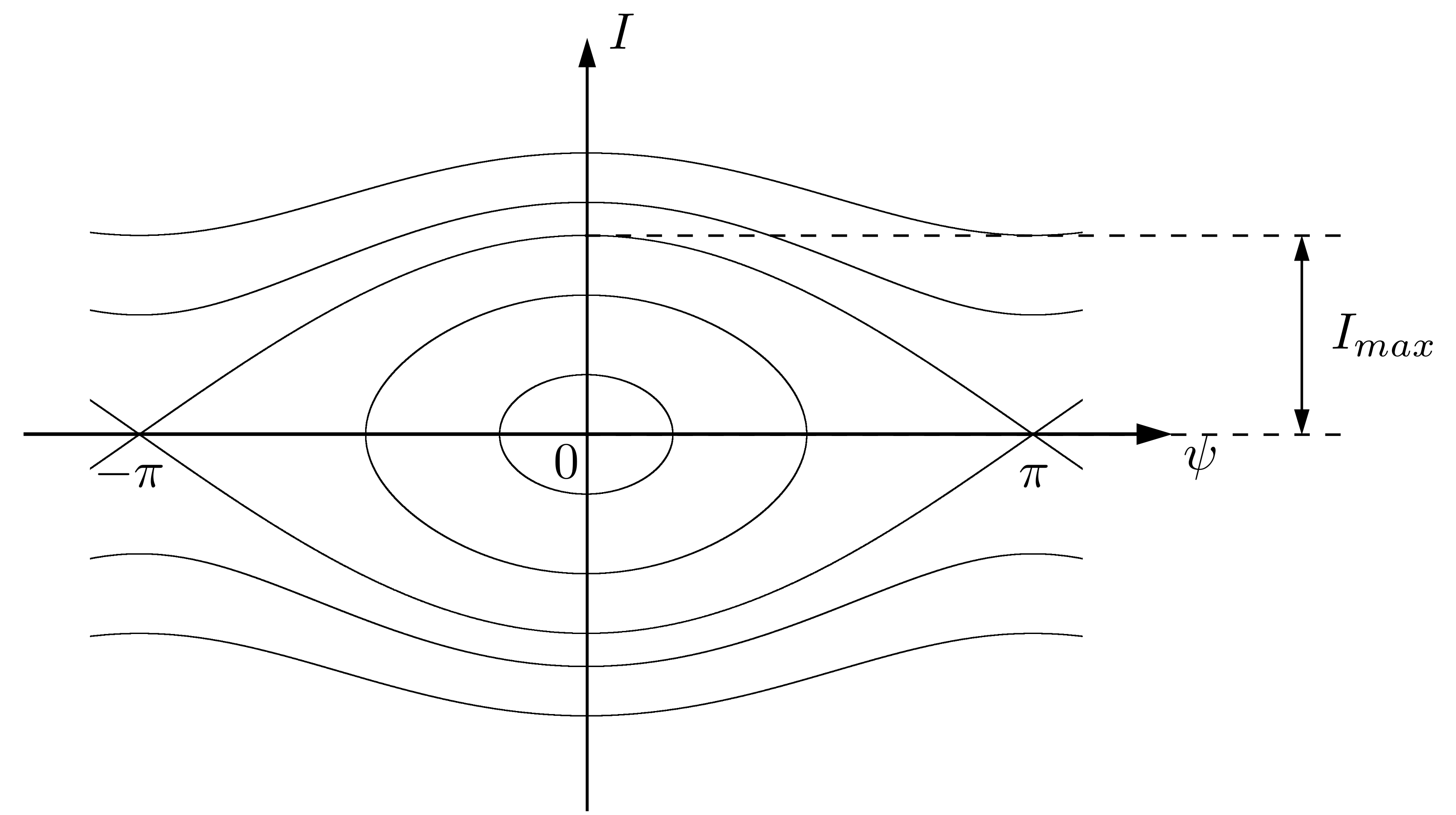}
		\caption{\label{fig_pend} Phase space for the pendulum described by the Hamiltonian function (\ref{8pendulo}). The trajectories can be closed curves around the elliptic point ($\psi=0,I=0$), indicating the oscillation around the fixed point, or the "open" curves that represent the rotation of the pendulum. The curve that connects the hyperbolic points $(\phi=\pm\pi, I=0$) is the separatrix.}
	\end{center}
\end{figure}

The phase trajectories described by the pendulum Hamiltonian are schematically represented in FIG. \ref{fig_pend}. We observe closed curves around the Poincaré-Birkhoff elliptic point $(I = 0, \phi = 0)$, with a separatrix connecting the hyperbolic points $(I = 0, \phi = \pm \pi)$. The half-width $I_{max}$ of this island structure corresponding to a $m/n$-resonance, is
	\begin{equation} 
		\label{8largura}
		I_{max} = 2  \, {\left\vert \frac{F}{G} \right\vert}^{1/2} = \frac{2a^2}{m}  \, \sqrt{\frac{\varepsilon\xi}{2\pi}}  \, {\left\{ 2 \left(1 - \frac{n q_a}{2m} \right) \right\}}^{m/4},
	\end{equation}
hence proportional to the factor $\sqrt{\varepsilon\xi} \ll 1$. By the same token, the oscillation frequency around the elliptic point is ${\left\vert FG \right\vert}^{1/2}$.

The pendulum Hamiltonian describing the resonance is integrable because we have averaged out the non-resonant terms in (\ref{8H1}). If we include these terms again, the system will become quasi-integrable and the pendulum separatrices will no longer join smoothly, but rather will present an infinite number of homoclinic and heteroclinic points. The dynamics near such points is known to be chaotic and, as a result, instead of separatrices, the islands will have a thin stochastic layer of chaotic motion \cite{lichtenberg}. As long as the intensity of perturbation is small enough, these locally chaotic layers do not connect themselves, preventing large-scale chaotic transport of field lines. If the limiter current, however, is larger than a critical value, the locally chaotic layers merge together forming a globally chaotic region, and allowing large-scale chaotic transport. Pendulum approximation have been used to estimate the islands width and to apply Chirikov criterion to find the critical perturbation amplitudes to create a chaotic layer in the plasma \cite{fernandes,karger2}. We remark that Escande and Momo have derived a similar formula for the island half-width without using Fourier components like $A_m(J^*)$, but rather using the magnetic flux through a ribbon which edges are lines passing by the elliptic and hyperbolic points in a given magnetic island \cite{momo}.
	
\subsection{Poincaré map of field lines}
	
The EML Hamiltonian (\ref{8H1}) exhibits an explicit dependency on the "time" $\phi$. For non-autonomous systems, the trajectories belong to an extended phase space, where the time is treated as a coordinate \cite{lichtenberg}. For the  quasi-integrable Hamiltonian system in (\ref{combined}), the solutions are in a three dimensional phase space and the flow is parameterized by the "time" $\phi$. In this way, a state is determined by three variables: $J$, $\theta$ and $\phi$.
	
Instead of studying the solution of the system in a three dimensional geometric space, we can reduce the dimensionality of the problem by the construction of a Poincaré surface. The Poincaré map is formed by the intersection of the solutions in the surface defined at a constant value of $\phi$. In this way, we have the values of $J$ and $\theta$ for when the magnetic field lines cross the surface, i.e., for each complete turn in the toroidal direction.
	
Associating the equations (\ref{8H0}), (\ref{8H1}) and (\ref{combined}), we obtain the Hamiltonian function for the magnetic field lines under the effect of EML,
\begin{eqnarray}
\begin{aligned}
	H(J, \theta, \phi)=\dfrac{2J}{q_a}&\left(1-\dfrac{J}{2a^2}\right) -\\
    &  \, \dfrac{\mu_0 I_L R_0}{\pi a^m B_0} \, (2J)^{m/2} \, \cos(m\theta) \, f(\phi). 
\end{aligned}
\end{eqnarray}
	
Defining a normalized action $I=J/(a^2/2)$, we can write a normalized Hamiltonian $\mathcal{H}=H/(a^2/2)$, given by 
	\begin{eqnarray}
		\mathcal{H}(I, \theta, \phi)=I\left(1-\dfrac{I}{4}\right)-2\, \varepsilon \, I^{m/2}  \, \cos(m\theta) \, f(\phi).
		\label{eq1}
	\end{eqnarray}
The only free control parameter in (\ref{eq1}) is the ratio $\varepsilon$ between the magnetic limiter current ($I_L$) and the total plasma current ($I_P$), for a fixed value of $m$.
	
From (\ref{eq1}), we obtain the Hamilton equations,
	\begin{eqnarray}
		\begin{aligned}
			\label{eq2}
			\dfrac{d\theta}{d\phi}=&\dfrac{\partial \mathcal{H}}{\partial I}=1-\dfrac{I}{2}-m\, \varepsilon \, I^{(m/2)-1}\, \cos(m\theta)\, f(\phi),\\
			\dfrac{d I}{d\phi}=&-\dfrac{\partial \mathcal{H}}{\partial \theta}=-2\, m\, \varepsilon \, I^{m/2} \, \sin (m\theta) \, f(\phi),
		\end{aligned}
	\end{eqnarray}
where $f(\phi)$ is the "time"-dependent factor given by (\ref{8fator})-(\ref{8fator1}), related to the perturbation created by a limiter ring of length $\ell$.
	
\begin{figure*}
	\begin{center}
		\includegraphics[width=0.8\textwidth]{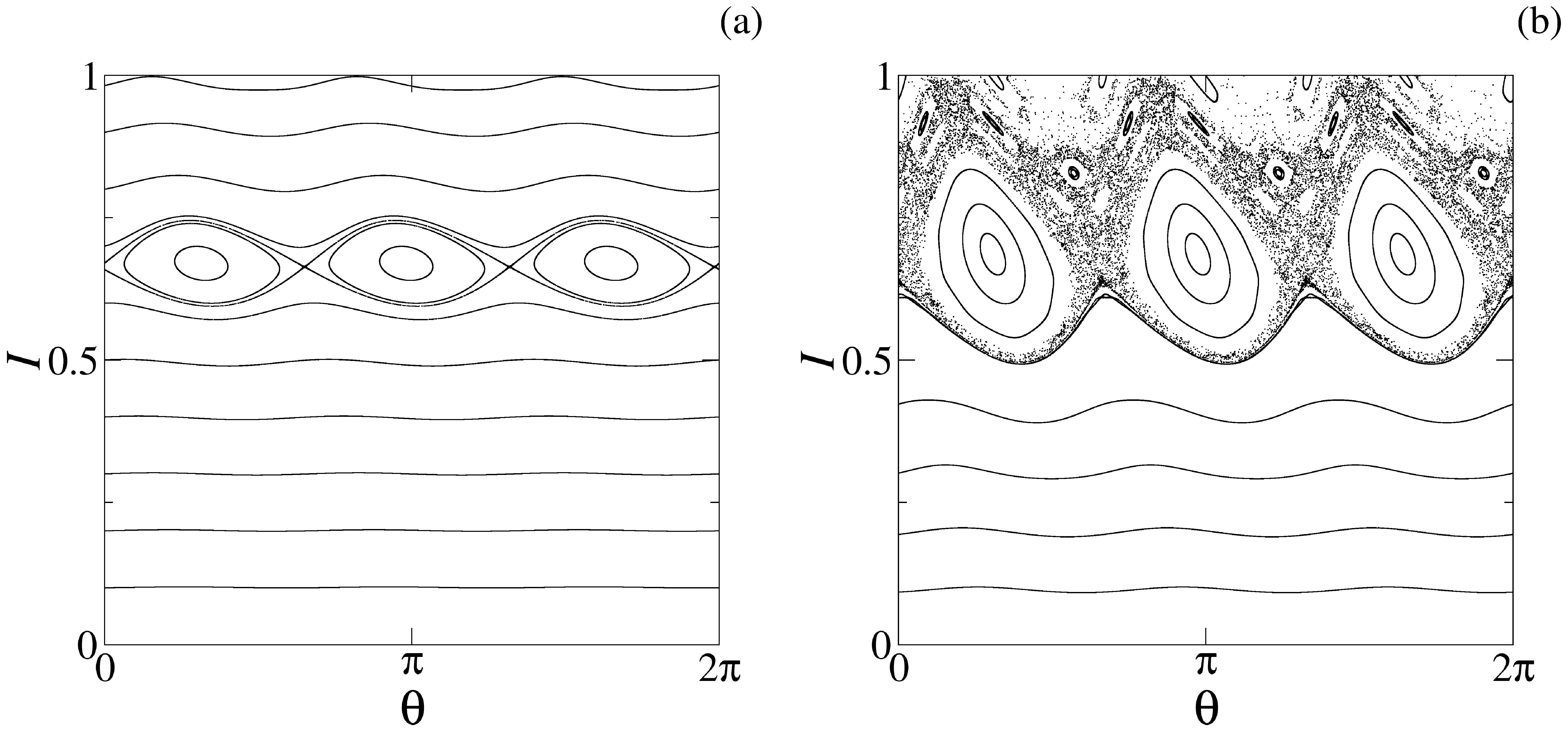}
		\caption{\label{PhaseSpace} Poincaré sections for the ergodic limiter systems, defined by (\ref{eq2}), with parameters $m=3$, $\xi=0.163$, for (a) $\varepsilon=0.025$ (small limiter current scenario), and (b) $\varepsilon=0.15$ (large limiter current scenario).}
	\end{center}
\end{figure*}

Integrating equations (\ref{eq2}), for initial conditions $(I(\phi=0),\theta(\phi=0))$, using a Symplectic Euler method and defining the Poincaré section at $\phi=0 \hspace{1em} (\text{mod} \hspace{0.5em} 2 \pi)$, we construct the Poincaré sections in FIG. \ref{PhaseSpace} (a) and (b), for a scenario of small ($\varepsilon=0.025$) and large ($\varepsilon=0.15$)  limiter currents, respectively.

For the Poincaré section in FIG. \ref{PhaseSpace} (a), the limiter current corresponds to $2.5\%$ of the total plasma current $I_L$. In this scenario, we observe regular solutions in most part of the space, represented by the rotational circles and the three islands (oscillatory circles). We observe a thin chaotic layer acting as a "separatrix" of the island. Since $\varepsilon$ is to small, the width of this chaotic layer is so tiny it resembles a separatrix curve, as the one presented in the pendulum phase space in FIG. \ref{fig_pend}. If the limiter current is increased until it corresponds to $15\%$ of the total plasma current (six times the current in FIG. \ref{PhaseSpace} (a)), we have the Poincaré section shown in FIG.\ref{PhaseSpace} (b). The second Poincaré section also show three islands, and the separatrix between them is replaced by a thick chaotic layer. The chaotic behavior emerges with the increase of the perturbation parameter $\varepsilon$, i.e., with the increase of the current in the ergodic limiter.

\begin{figure*}
 \begin{center}
	\includegraphics[width=0.6\textwidth]{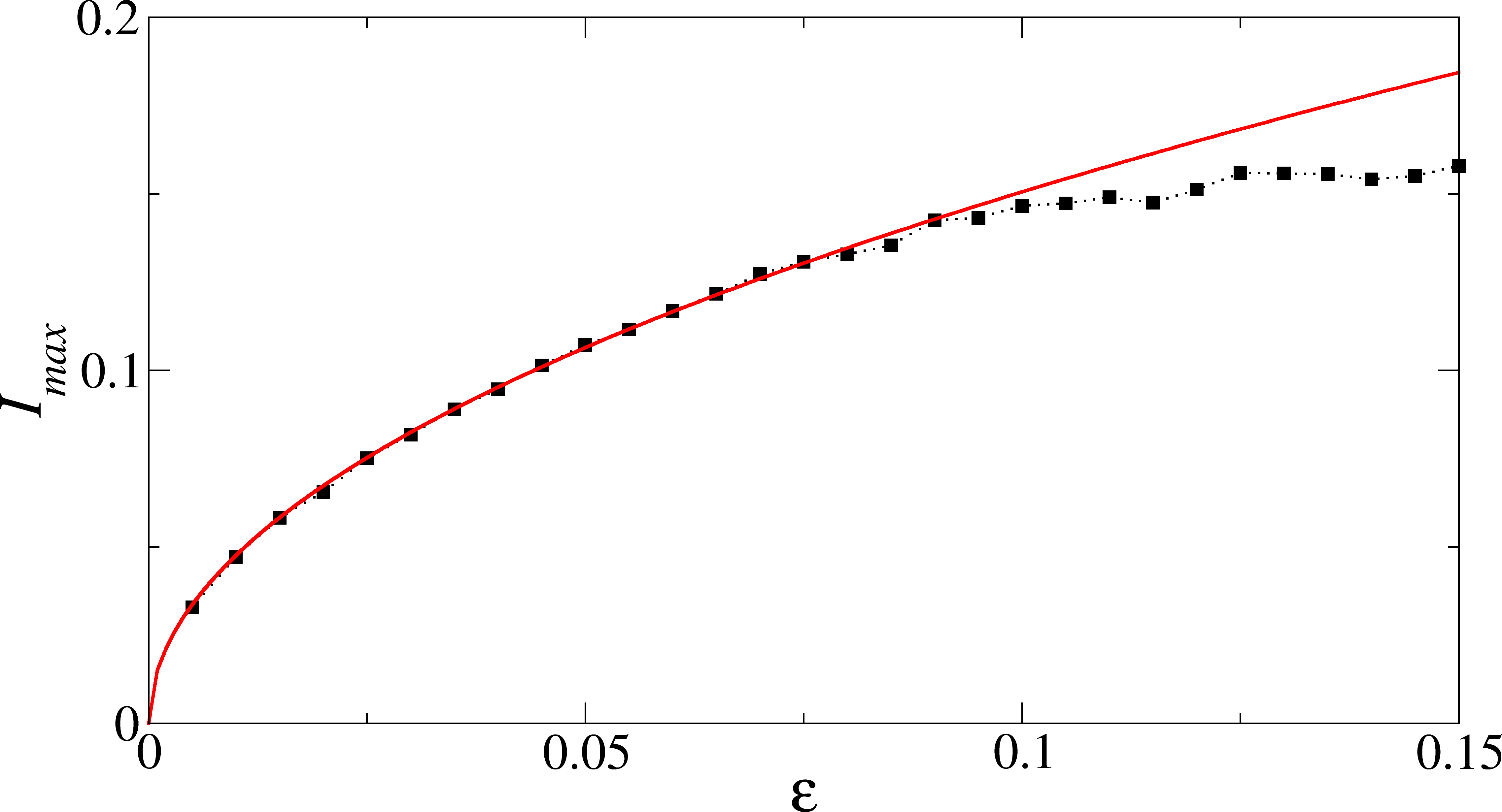}
	\caption{\label{eps} Half-width of the islands in the phase space for different values of $\varepsilon$. The value of $I_{max}$, indicated by the black squares, was obtained analysing the islands for the Hamiltonian system  (\ref{eq1}) for each $\varepsilon$, with $m=3$ and $\xi=0.163$. The red curve is obtained by the equation (\ref{8largura}) multiplied by a scale factor. For smaller values of $\varepsilon$, the red curve agrees with the black points, i.e., $I_{max}$ follows the relation $I_{max}\propto \varepsilon^{1/2}$ given by the pendulum approximation for $\varepsilon \lesssim 0.1$.}
\end{center}
\end{figure*}

Finally, we would like to estimate until what value of $\varepsilon$ the equation (\ref{8largura}) is a good approximation for the half-width of the islands in the phase space. We numerically solve the system (\ref{eq2}), construct the Poincaré section and compare the half-width of the islands for each $\varepsilon$ with the value predicted by Equation (\ref{8largura}). The results are presented in FIG. \ref{eps} and we observe that the pendulum approximation is valid, for the half-width of the islands in the phase space, until $\varepsilon \approx0.1$. For higher values of $\varepsilon$, the value of $I_{max}$ does not increase at the same rate proportional to $\varepsilon^{1/2}$. For $\varepsilon>0.1$, we observe small increases and decreases in $I_{max}$,  represents the enlargement and the following destruction of the island..
	
\section{Magnetic flux coordinates}
	
Magnetic flux coordinates are often used both in theoretical and computational studies of MHD equilibria in toroidal plasma devices.\cite{shohet}. They are denoted by $(\psi,\theta,\zeta)$, where $\psi$ is a magnetic surface label, whereas $\theta$ and $\zeta$ are angle-like variables, often called poloidal and toroidal angles, although they do not have a direct geometrical meaning as the angles $(\theta,\phi)$ introduced in section IV. The contravariant coordinates are thus
	\begin{equation}
		\label{fluxcoord}
		x^1 = \theta, \qquad x^2 = \psi, \qquad x^3 = \zeta,
	\end{equation}
such that $\theta$ and $\zeta$ increase by $2\pi$ after a complete turn around the torus.
	
The equilibrium MHD condition \cite{miyamoto}
	\begin{equation}
		\label{equil}
		\nabla p = {\bf j}\times{\bf B}
	\end{equation}
states that the expanding tendency caused by plasma pressure is counterbalanced by a magnetic force produced by a current density ${\bf j}$ interacting with the resultant magnetic field ${\bf B}$. 
Dotting (\ref{equil}) with ${\bf B}$ results that 
	\begin{equation}
		\label{fluxsurf}
		{\bf B}\cdot\nabla p = 0,
	\end{equation}
such that magnetic field lines lie on constant pressure surfaces, also called flux surfaces or magnetic surfaces. In general, a quantity $\psi$ is a magnetic surface label if $\psi = const.$ on all its points, 
	\begin{equation}
		\label{fluxconst}
		{\bf B}\cdot \nabla\psi = 0.    
	\end{equation}
In other words, a magnetic surface is a coordinate surface for which $x^2 = \psi = const.$. Moreover, the magnetic axis - which is a $x^3$-coordinate curve - is a degenerate magnetic surface of zero volume, with $\psi = 0$ on each of its points. 
	
Various physical quantities can play the role of a magnetic surface label, among them the pressure itself, but also the volume enclosed by a magnetic surface, for example. However, in the specific case of flux coordinates, $\psi$ is chosen to be proportional to the toroidal flux, i.e. to the magnetic flux enclosed by a magnetic surface:
	\begin{equation}
		\label{torflux}
		\psi = \frac{1}{2\pi} \int_{S_t} {\bf B}\cdot{\bf dS}^{(3)} =   \frac{1}{2\pi} \int_{S_t} \nabla\times{\bf A}\cdot{\bf dS}^{(3)},
	\end{equation}
where the vectorial area element perpendicular to a coordinate surface $x^3 = const.$ is 
	\begin{equation}
		\label{ds3}
		{\bf dS}^{(3)} = \sqrt{g} dx^1 dx^2 {\hat{\bf e}}^3, 
	\end{equation}
and $S_t$ is the cross-section of the magnetic surface with the plane $\zeta = const.$
	
Applying Stokes theorem in (\ref{torflux}) gives
	\begin{equation}
		\label{torflux1}
		\psi = \frac{1}{2\pi} \oint_{C_t} {\bf A}\cdot{\bf dl}, 
	\end{equation}
where $C_t$ is the boundary of the surface $S_t$, with the line element 
	\begin{equation}
		\label{dele}
		{\bf dl} = dx^1 {\hat{\bf e}}_1,
	\end{equation}
such that the path integral corresponds to a short turn along the poloidal angle $\theta$, 
	\begin{equation}
		\label{torflux2}
		\psi = \frac{1}{2\pi} \int_0^{2\pi} A_1 dx^1. 
	\end{equation}
	
If $A_1$ does not depend on $x^1$, then the canonical momentum is the toroidal flux itself: 
	\begin{equation}
		\label{canonmom}
		p = A_1 = \psi,
	\end{equation}
canonically conjugated to the coordinate
	\begin{equation}
		\label{canoncor}
		q = x^1 = \theta,
	\end{equation}
and with the time-like variable equal to the toroidal angle
	\begin{equation}
		\label{timecor}
		t = x^3 = \zeta,
	\end{equation}
which is an ignorable coordinate for toroidal magnetic surfaces.
	
Now consider the magnetic flux through a ribbon-like surface that extends from the magnetic axis to the magnetic surface, being a coordinate surface $x^1 = const.$. The poloidal flux is proportional to the magnetic flux through this surface $S_p$
	\begin{equation}
		\label{polflux}
		\alpha = \frac{1}{2\pi} \int_{S_p} {\bf B}\cdot{\bf dS}^{(1)} =   \frac{1}{2\pi} \int_{S_p} \nabla\times{\bf A}\cdot{\bf dS}^{(1)},
	\end{equation}
where 
	\begin{equation}
		\label{ds1}
		{\bf dS}^{(1)} = \sqrt{g} dx^2 dx^3 {\hat{\bf e}}^1.
	\end{equation}
is the vectorial area element perpendicular to a coordinate surface $x^1 = const.$. By Stokes' theorem it results 
	\begin{equation}
		\label{polflux1}
		\alpha = \frac{1}{2\pi} \oint_{C_p} {\bf A}\cdot{\bf dl}, 
	\end{equation}
where $C_t$ is the boundary $S_t$, with the line element 
	\begin{equation}
		\label{dele1}
		{\bf dl} = - dx^3 {\hat{\bf e}}_3,
	\end{equation}
and the integral amounts to a long turn along the toroidal angle $\theta$, 
	\begin{equation}
		\label{polflux2}
		\alpha = - \frac{1}{2\pi} \int_0^{2\pi} A_3 dx^3. 
	\end{equation}
	
If $A_3$ does not depend on $x^3$, then the field line Hamiltonian is the poloidal flux, 
	\begin{equation}
		\label{canonham}
		H = - A_3 = \alpha.
	\end{equation}
With these associations the magnetic field line equations are equivalent to Hamilton´s equations
	\begin{eqnarray}
		\label{e1}
		\frac{dq}{dt} = \frac{\partial H}{\partial p}, \qquad & \Rightarrow & \qquad 
		\frac{d\theta}{d\zeta} = \frac{\partial\alpha}{\partial\psi}, \\
		\label{e2}
		\frac{dp}{dt} =  - \frac{\partial H}{\partial q}, \qquad & \Rightarrow & \qquad 
		\frac{d\psi}{d\zeta} = - \frac{\partial\alpha}{\partial\theta}.
	\end{eqnarray}
	
\subsection{Clebsch representation}
	
Since ${\hat{\bf e}}^i = \nabla x^i$ are the contravariant basis vectors, according to (\ref{fluxcoord}) we can write the vector potential in magnetic flux coordinates as
	\begin{equation}
		\label{Amag}
		{\bf A} = A_\theta \nabla\theta + A_\psi \nabla\psi + A_\zeta \nabla\zeta.
	\end{equation}
Let us define a scalar function $G$ by the condition,
	\begin{equation}
		\label{Gdef}
		\frac{\partial G}{\partial\psi} = A_\psi,
	\end{equation}
such that its gradient is 
	\begin{equation}
		\label{gradG}
		\nabla G = \frac{\partial G}{\partial\theta} \nabla\theta  + A_\psi \nabla\psi + \frac{\partial G}{\partial\zeta} \nabla\zeta.
	\end{equation}
	
Subtracting (\ref{gradG}) from (\ref{Amag}) we have
	\begin{equation}
		\label{A2}
		{\bf A} = \nabla G + \left( A_\theta - \frac{\partial G}{\partial\theta} \right) \nabla\theta + \left( A_\zeta - \frac{\partial G}{\partial\zeta} \right) \nabla\zeta
	\end{equation}
	
We define the toroidal and magnetic fluxes in terms of the derivatives of the scalar function introduced in (\ref{Gdef})
	\begin{eqnarray}
		\label{psidef}
		\psi & = A_\theta - \dfrac{\partial G}{\partial\theta} \\
		\label{alphadef}
		\alpha & = - A_\zeta + \dfrac{\partial G}{\partial\zeta}.
	\end{eqnarray}
Note that this amounts to choose a determined gauge. Substituting both expressions in (\ref{A2}) the vector potential reads
	\begin{equation}
		\label{A}
		{\bf A} = \nabla G + \psi\nabla\theta - \alpha\nabla\zeta.
	\end{equation}
	
Taking the rotational of this expression and using standard vector identities we obtain the magnetic field in terms of the flux coordinates in the form
	\begin{equation}
		\label{clebsch0}
		{\bf B} = \nabla\psi \times \nabla\theta - \nabla\alpha \times \nabla\zeta,
	\end{equation}
also known as Clebsch representation \cite{shohet}. This is the most general representation of a magnetic field satisfying simultaneously the conditions ${\bf B}\cdot\nabla\psi = 0$ and $\nabla\cdot{\bf B} = 0$.
	
Observe that, using flux coordinates, it is possible to express the safety factor, which is a surface quantity, as the ratio between these fluxes:
	\begin{equation}
		\label{sf}
		q(\psi) = \frac{d\zeta}{d\theta} = \frac{\partial\psi}{\partial\alpha},
	\end{equation}
provided we know the function $\alpha = \alpha(\psi)$. In this case 
	\begin{equation}
		\nabla\alpha(\psi) = \frac{\partial\alpha}{\partial\psi} \nabla \psi = \frac{1}{q(\psi)} \nabla\psi = \frac{\iota(\psi)}{2\pi} \nabla\psi,
	\end{equation}
and the Clebsch representation (\ref{clebsch0}) reads
	\begin{equation}
		\label{clebsch}
		{\bf B} = \nabla\psi \times \nabla\theta - \iota(\psi) \nabla\psi \times \nabla\zeta,
	\end{equation}
	where the factor $2 \pi$ is absorbed in the rotational transform $\iota$.
	
\subsection{Exploitation of the variational principle}
	
Let us exploit the variational principle for field lines
	\begin{equation}
		\label{hamf}
		\delta \int_{{\bf r}_1}^{{\bf r}_2} {\bf A}\cdot{\bf dr} = \int_1^2 \delta({\bf A}\cdot{\bf dr}) = 0,
	\end{equation}
in order to obtain the Euler-Lagrange equations, that correspond to the magnetic field line equations. Using (\ref{A}) we have the variation,
	\begin{equation}
		\label{A1}
		{\bf A}\cdot{\bf dr} = \left\{ \frac{dG}{d\zeta} + \psi \frac{d\theta}{d\zeta} - \alpha \right\} d\zeta.
	\end{equation}
	
Inserting 
	\[
	\delta \alpha = \frac{\partial\alpha}{\partial\psi} \delta\psi + \frac{\partial\alpha}{\partial\theta} \delta\theta,
	\]
in (\ref{A1}) it follows that (\ref{hamf}) becomes
\begin{widetext}
\begin{equation}
	\label{A3}
	\int_1^2 \left\{
	\left( \frac{d\theta}{d\zeta} - \frac{\partial\alpha}{\partial\psi} \right) \delta\psi - 
	\left( \frac{d\psi}{d\zeta} + \frac{\partial\alpha}{\partial\theta} \right) \delta\theta + \frac{d}{d\zeta} (\delta G + \psi \delta\theta)  \right\} d\zeta = 0.
\end{equation}
\end{widetext}
	
The third term inside the braces vanishes because
	\[
	\int_1^2 \frac{d}{d\zeta} (\delta G + \psi \delta\theta) d\zeta = 
	{(\delta G + \psi \delta\theta)}_2 - {(\delta G + \psi \delta\theta)}_1 = 0,
	\]
since $1$ and $2$ are fixed points. Hence 
	\begin{equation}
		\label{A3}
		\int_1^2 \left\{
		\left( \frac{d\theta}{d\zeta} - \frac{\partial\alpha}{\partial\psi} \right) \delta\psi - 
		\left( \frac{d\psi}{d\zeta} + \frac{\partial\alpha}{\partial\theta} \right) \delta\theta \right\} d\zeta = 0,
	\end{equation}
that holds for arbitrary variations in $\psi$ and $\theta$ if the coefficients vanish identically, giving 
	\begin{eqnarray}
		\label{e11}
		\frac{d\theta}{d\zeta} = \frac{\partial\alpha}{\partial\psi}, \\
		\label{e21}
		\frac{d\psi}{d\zeta} = - \frac{\partial\alpha}{\partial\theta},
	\end{eqnarray}
which are the magnetic field line equations (\ref{e1})-(\ref{e2}), written in canonical form. Observe that this results independs on the gauge used, since the term $G$ disappears during the calculation. 
	
\subsection{Tokamap Hamiltonian}
	
Let us first consider, as an example, the integrable equilibrium tokamak magnetic field, for which the Hamiltonian $\alpha$ depends only on the canonical momentum $\psi$. In this case we can rewrite Hamilton´s equations (\ref{e11}) and (\ref{e21}) in the following form
	\begin{eqnarray}
		\label{e11a}
		\frac{d\theta}{d\zeta} & = & \frac{1}{q(\psi)}, \\
		\label{e21a}
		\frac{d\psi}{d\zeta} & = & 0,
	\end{eqnarray}
where $q(\psi) = \partial\alpha_0(\psi)/\partial\psi$ is the safety factor of the unperturbed magnetic surfaces (the subscript in $\alpha$ stands for this fact). In this case we identify $\psi$ as an action variable, $\theta$ being its conjugate angle. Since $\psi$ is a constant of motion (parameterized by the timelike variable $\zeta$) the equilibrium consists on nested tori, that can be rational or irrational according to the corresponding value of $q(\psi)$. 
	
A magnetostatic non-symmetric perturbation can be represented, as in the cylindrical case, by a term $\delta\alpha(\psi,\theta,\zeta)$ in the field line Hamiltonian, which now reads
	\begin{equation}
		\label{deltalfa}
		\alpha = \alpha_0(\psi) + K \,  \delta\alpha(\psi,\theta,\zeta),
	\end{equation}
where $K > 0$ is a parameter which represents the strength of the perturbation, with respect to the equilibrium. The corresponding Hamilton´s equations
	\begin{eqnarray}
		\label{e11b}
		\frac{d\theta}{d\zeta} & = & \frac{1}{q(\psi)} + K \frac{\partial\delta\alpha(\psi,\theta,\zeta)}{\partial\psi}, \\
		\label{e21b}
		\frac{d\psi}{d\zeta} & = & - K \frac{\partial\delta\alpha(\psi,\theta,\zeta)}{\partial\theta},
	\end{eqnarray}
can, in principle, be integrated with respect to the timelike variable $\zeta$. A Poincaré map is obtained by sampling the values of $(\psi,\theta)$ at fixed intervals of $\zeta$. If, as it is often assumed, we sample variables after a complete toroidal turn, then $(\psi_n,\theta_n)$ are the values of the action and angle variables at the $n$th piercing of the magnetic field line with a plane $\zeta = const.$
	
Since the coordinates of each piercing are unique functions of the coordinates of the previous one, we are able obtain a two-dimensional map in the general form
	\begin{eqnarray}
		\label{map1}
		\psi_{n+1} & = & {\cal A}(\psi_n,\theta_n), \\
		\label{map2}
		\theta_{n+1} & = & {\cal B}(\psi_n,\theta_n),
	\end{eqnarray}
where $n = 0, 1, 2, \ldots$ can be interpreted as a discrete timelike variable, and ${\cal A}$ and ${\cal B}$ are functions that can be obtained analytically in some special cases.
	
It is known, from Hamiltonian dynamics, that the above map represents a canonical transformation $(\psi_n,\theta_n) \rightarrow (\psi_{n+1},\theta_{n+1})$, corresponding to a generating function of the second kind, written as\cite{lichtenberg},
	\begin{equation}
		\label{gf}
		{\cal F}(\psi_{n+1},\theta_n) = \psi_{n+1} \theta_n + {\cal F}_0(\psi_{n+1}) + K \delta{\cal F}(\psi_{n+1},\theta_n),
	\end{equation}
where the first term generates the identity transformation, and the second and third terms are related to the equilibrium and perturbation, respectively. 
	
The equations for this canonical transformation are 
	\begin{eqnarray}
		\label{ct1}
		\psi_n & = & \frac{\partial{\cal F}}{\partial\theta_n} = \psi_{n+1} + K \frac{\partial\delta{\cal F}}{\partial\theta_n}, \\
		\label{ct2}
		\theta_{n+1} & = & \frac{\partial{\cal F}}{\partial\psi_{n+1}} = \theta_n + \frac{d{\cal F}_0}{d\psi_{n+1}} + 
		K \frac{\partial\delta{\cal F}}{\partial\psi_{n+1}}.
	\end{eqnarray}
	
Starting again from the unperturbed case $(K = 0)$ we have that 
	\begin{eqnarray}
		\label{ct1}
		\psi_n & = & \psi_{n+1} \\
		\label{ct2}
		\theta_{n+1} & = & \theta_n + \frac{d{\cal F}_0}{d\psi_{n+1}},
	\end{eqnarray}
which is just the solution of the equations (\ref{e11a}) and (\ref{e21a}), provided we make the identification
	\begin{equation}
		\label{wdef}
		\frac{1}{q(\psi)} = \frac{d{\cal F}_0}{d\psi}.
	\end{equation}
	
For considering the perturbed case it is useful to define the following functions
	\begin{eqnarray}
		\label{hdef}
		h(\psi_{n+1},\theta_n) & = & -  \frac{\partial\delta{\cal F}}{\partial\theta_n}, \\
		\label{jdef}
		j(\psi_{n+1},\theta_n) & = & \frac{\partial\delta{\cal F}}{\partial\psi_{n+1}}, 
	\end{eqnarray}
thus satisfying the condition
	\begin{equation}
		\label{condi}
		\frac{\partial h}{\partial\psi_{n+1}} + \frac{\partial j}{\partial \theta_n} = 0,
	\end{equation}
in such a way that the map equations are
	\begin{eqnarray}
		\label{m1}
		\psi_{n+1} & = & \psi_n + K h(\psi_{n+1},\theta_n), \\
		\label{m2}
		\theta_{n+1} & = & \theta_n + \frac{1}{q(\psi_{n+1})} + K j(\psi_{n+1},\theta_n).
	\end{eqnarray}
	
Another consequence of the above map being a canonical transformation is that it preserves areas in the Poincaré surface of section: $d\psi_{n+1} d\theta_{n+1} = d\psi_n d\theta_n$. This implies that the Jacobian of the transformation has an absolute value equal to the unity, i.e. $|{\cal J}| = 1$, where 
	\begin{equation}
		\label{jacobian}
		{\cal J} = \left\vert
		\begin{matrix}
			\partial \psi_{n+1}/\partial \psi_n & \partial\psi_{n+1}/\partial\theta_n \\
			\partial \theta_{n+1}/\partial \psi_n & \partial\theta_{n+1}/\partial\theta_n 
		\end{matrix}    
		\right\vert.
	\end{equation}
In fact, the map (\ref{m1})-(\ref{m2}) is area preserving, provided (\ref{condi}) is fulfilled. 
	
In order to put these equations into the form (\ref{map1})-(\ref{map2}) it is necessary to solve them first for $\psi_{n+1}$. Although this can be done analytically in some cases, it is always possible to use root-finding methods to do so numerically. Another important point, emphasized by Balescu and coauthors, is that the field line map must have two properties: (i) if $\psi_0 > 0$ then $\psi_n > 0$, for all values of $n$; (ii) if $\psi_0 = 0$ then $\psi_n \ge 0$ for all $n$ \cite{balescu}. The former property comes from the definition of the coordinate $\psi$, which must be a definite positive number, whereas the latter is related to the fact that $\psi = 0$ stands for the magnetic axis (which is a degenerate magnetic surface).
	
Balescu proposed a map (called tokamap) that satisfies both requirements, namely \cite{balescu}
	\begin{eqnarray}
		\label{t1}
		\psi_{n+1} & = & \psi_n - \frac{K}{2\pi} \frac{\psi_{n+1}}{1+\psi_{n+1}} \sin(2\pi\theta_n), \\
		\label{t2}
		\theta_{n+1} & = & \theta_n + \frac{1}{q(\psi_{n+1})} - \frac{K}{{(2\pi)}^2} \frac{\cos(2\pi\theta_n)}{{(1+\psi_{n+1})}^2} .
	\end{eqnarray}
It is actually possible to analytically solve (\ref{t1}) for $\psi_{n+1}$ but there are two solutions for a given $(\psi_n,\theta_n)$. We can preserve uniqueness by choosing the positive root, viz.
\begin{equation}
\label{pr}
	\psi_{n+1} = \frac{1}{2} \left\{
	P(\psi_n,\theta_n) + \sqrt{{[P(\psi_n,\theta_n)]}^2 + 4 \psi_n}
	\right\},
\end{equation}
where 
\begin{equation}
	\label{pdef}
	P(\psi,\theta) = \psi - 1 - \frac{K}{2\pi} \sin(2\pi\theta).
\end{equation}
	
\begin{figure*}
	\begin{center}
		\includegraphics[width=0.85\textwidth]{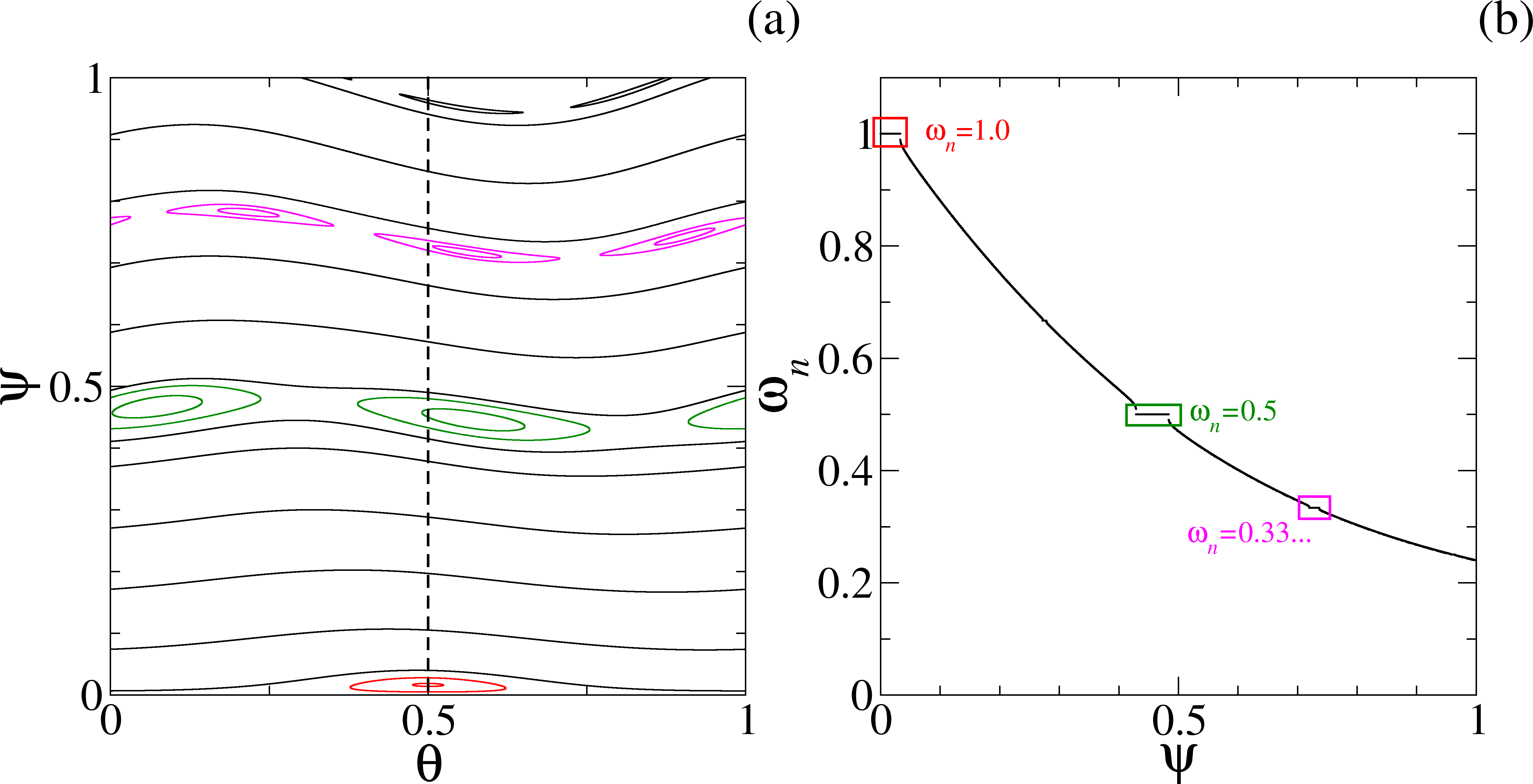}
		\caption{\label{wn_tm1} Solutions of the tokamap for $K=1$. For the phase space in (a), we only observe periodic and quasi-periodic solutions and their respective winding numbers are showed in the profile in (b). The highlight winding number plateaus in (b) correspond to the colored islands with the same color in (a).}
	\end{center}
\end{figure*}

Balescu and co-workers \cite{balescu} have studied the properties of the tokamap for the following choice of safety factor 
	\begin{equation}
		\label{tokaq}
		q(\psi) = \frac{4 q_0}{(2-\psi)(2-2\psi+\psi^2)},
	\end{equation}
where $q_0$ is the safety factor at the magnetic axis $\psi = 0$. For numerical simulations it has been conventioned that $\psi = 1$ is the position of the tokamak wall, such that $0 \le \psi \le 1$ is the physical range of this variable. The safety factor increases monotonically to the tokamak wall, where it is $q(\psi = 1) = 4$. Since this function is monotonically increasing, the corresponding tokamap (\ref{t1})-(\ref{t2}) satisfies the twist property.  As a consequence of the monotonic increase of the safety factor, the winding number profile presents a monotonic behavior in $\psi$. The winding number for a solution of a system that exhibits a periodicity in the $\theta$ variable and is defined by the limit,
	\begin{eqnarray}
		\omega_n=\lim_{n\to \infty} \dfrac{\theta_n-\theta_0}{n}
		\label{wn}
	\end{eqnarray}
that converges for periodic and quasi-periodic solutions, while is not defined for chaotic trajectories. For $K=1$, we have the Poincaré section and the winding number profile, calculated in $\theta=0.5$ for a final iteration time of $10^6$ iterations, showed in FIG. \ref{wn_tm1}.
	
We observe that the tokamap exhibits mainly periodic and quasi-periodic solutions for $K=1$, indicated by the existence of only islands and rotational circles in the Poincaré section of FIG. \ref{wn_tm1} (a). Like in the previous section, if the perturbation strength is to small, the size of a chaotic layer in the neighbourhood of an island is so tiny that can be revealed only by magnifications of the Poincaré section. From the winding number profile calculated in $\theta=0.5$ [FIG. \ref{wn_tm1} (b)] we observe a defined $\omega_n$ for almost every value of $\psi$ and the profile monotonically decreases. The possible exceptions consist of tiny intervals for which the orbit is chaotic. Each island is represented by a plateau in the profile while the regions of rotational circles exhibit a different value of $\omega$ for each solution.
	
In FIG. \ref{wn_tm1} (b), we choose three different plateaus and highlight them with the colored rectangles. The correspondent island of the plateau is show with the same color in the Poincaré section of FIG. \ref{wn_tm1} (a). From the values of $\omega_n$, we identified a direct relation with the period of the islands. The winding number is related to the period $\tau$ of the island by $\omega_n=1/\tau$. For example, the red island of period is $1$ presents a winding number equal to $\omega_n=1/1$. The green and pink islands presents winding numbers equal to $\omega_n=0.5=1/2$ and $\omega_n=0.33...=1/3$, respectively. These periodic islands are on rational tori, since we can write their frequencies as a ration between two integer numbers.
	
\begin{figure*}
\begin{center}
	\includegraphics[width=0.85\textwidth]{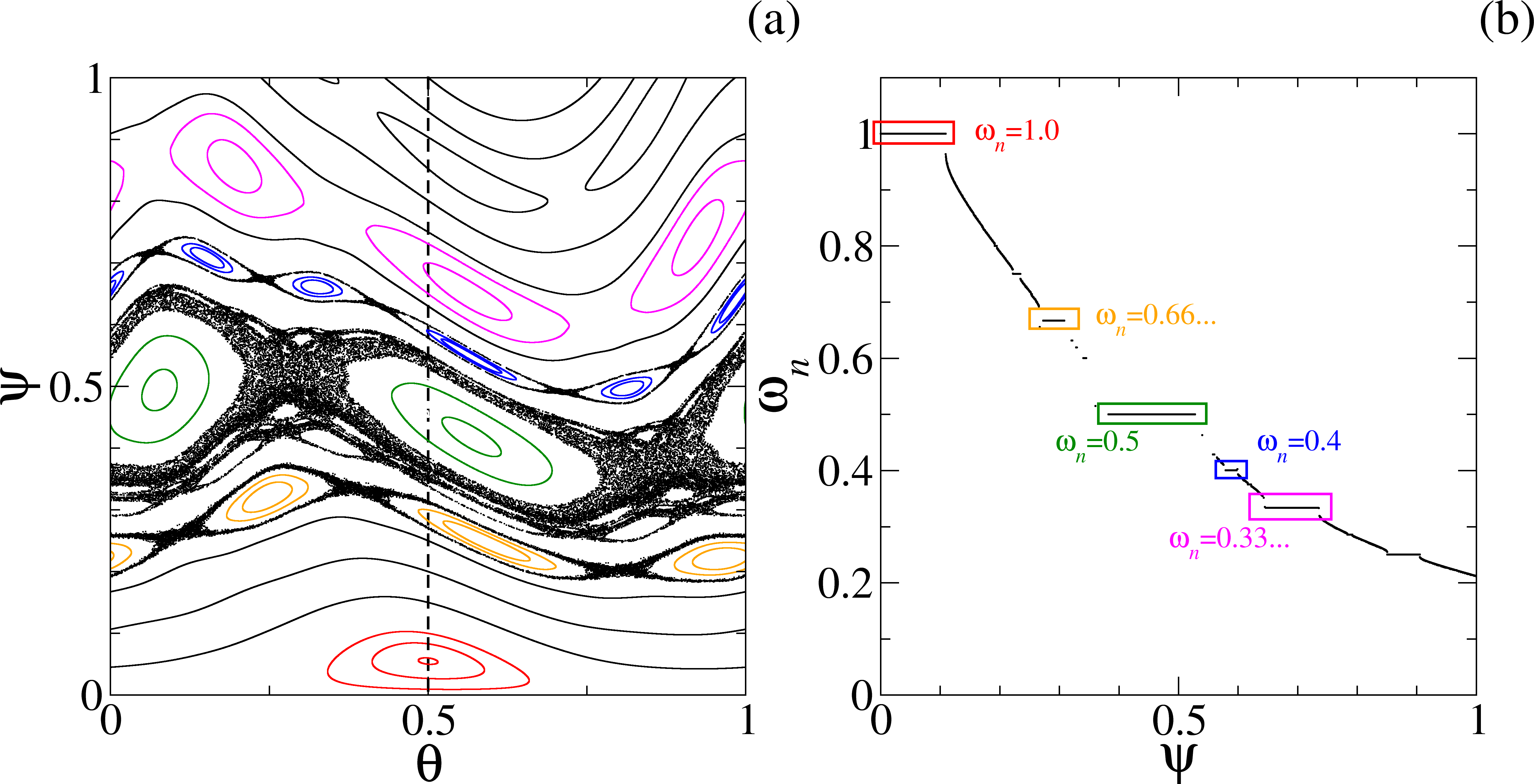}
	\caption{\label{wn_tm2} The tokamap for $K=3.5$. The phase space in (a) exhibit, periodic and quasi-periodic solutions as a chaotic behavior represented by the sparse points around the green and orange islands. The winding number profile calculated in $\theta=0.5$ is showed in (b).}
\end{center}
\end{figure*}

Increasing the perturbation parameter to $K=3.5$, we have the Poincaré section and the winding number profile showed in FIG. \ref{wn_tm2}, where  we observe that some regular solutions are replaced by stochastic layers, represented by the chaotic seas around the green and the orange lines. The chaotic behavior is restricted, and the chaotic regions are not connected. If we increase $K$, these chaotic regions will eventually enlarge and merge together into a single area-filling chaotic orbit. Following the same methodology as for the FIG. \ref{wn_tm1}, we computed the winding number profile, at $\theta=0.5$, and highlight the plateaus of constant $\omega_n$ values. Again, we observed the directed relation between the winding number value and the period $\tau$ of the islands. The red, green and pink islands of FIG. \ref{wn_tm1} are also seen here, with $\omega_n=1/1=1$, $\omega_n=1/2=0.5$ and $\omega_n=1/3=0.33...$, respectively. We also highlight two other chains of islands, orange and blue, with $\omega_n=2/3=0.66...$ and $\omega_n=2/5=0.4$, respectively. For these last two islands, we have that the trajectory always "jumps" one island during the time evolution, \textit{i.e.}, if we choose an initial condition in the blue island close to $\theta=0.5$ (the third island counting from left to right), the next point will be in the fifth island, the second iteration will be in the second island, an so on. The chaotic regions are represented by the "gaps" in the winding number profile, since the limit (\ref{wn}) does not converge.
 
The Tokamap have been used to interpret the particle escape to the wall in Textor Tokamak. In particlular the theoretically  obtained fractal distribution of field lines at the plasma edge  is similar to the one measured in this tokamak \cite{abdullaev, jakubowski}.

\subsection{Analysis of the revtokamap}
	
In a subsequent paper, Balescu has used another choice for the safety factor, namely \cite{balescu2,schelin}
	%\begin{equation}
	%    \label{tokaqr}
	%    q(\psi) = \frac{q_0}{1 - a{(c\psi - 1)}^2},
	%\end{equation}
	\begin{eqnarray}
		q(\psi)=\dfrac{q_m}{1-a(\psi-\psi_m)^2}
		\label{tokaqr}
	\end{eqnarray}
which is a non-monotonic function of $\psi$ and the corresponding map does not satisfy the twist condition ($\partial \theta_{n+1}/\partial \psi_n \ne 0$)\cite{morrison}. As a consequence of the violation of the twist condition, the winding number profile presents a non-monotonic behavior. It has been called revtokamap, since it describes a profile with reversed shear, with an extreme (shearless point). The minimum $\psi_m$ of the profile (\ref{tokaqr}) is given by,
	\begin{eqnarray}
		\psi_m=\left(1+\sqrt{\dfrac{1-q_m/q_1}{1-q_m/q_0}}\right)^{-1}.
	\end{eqnarray}
The parameters $q_0$, $q_m$ and $q_1$ are chosen to reproduce approximately experimental data, and $a$ is defined as $a=(1-q_m/q_0)/\psi_m^2$.
	
The violation of the twist property in the Poincaré section brings consequences to the solutions of the map. Firstly, the extremum point in the shear correspond to the extremum point in the winding number profile. This extremum point belongs to the shearless curve in the phase space. Secondly, since the map is nontwist, two solutions can be isochronous, \textit{i.e.}, two distinct solutions present same period and winding number.
	
\begin{figure*}
\begin{center}
	\includegraphics[width=0.85\textwidth]{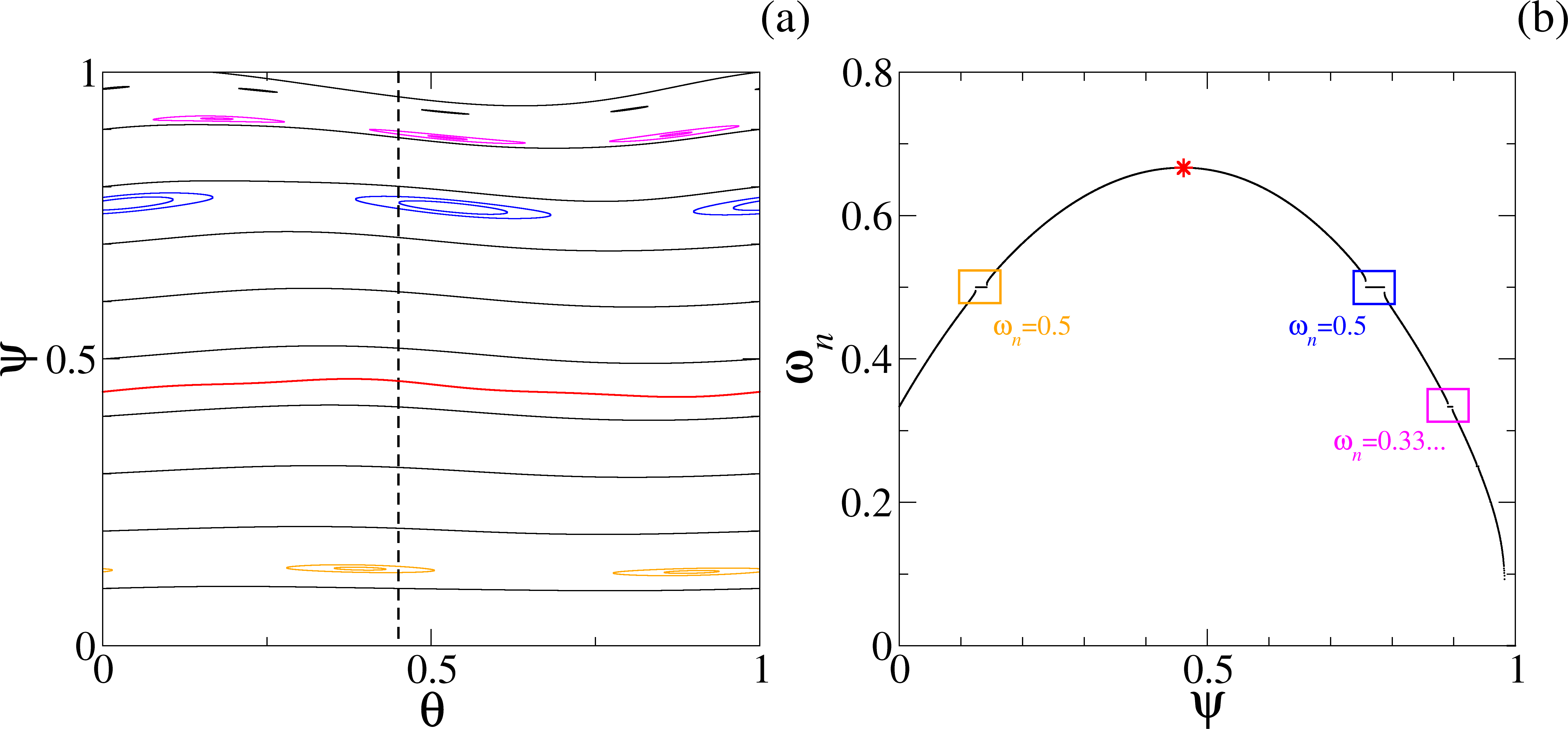}
	\caption{\label{rtm1} The Revtokamap for a lower perturbation parameter ($K=0.5$). The (a) extreme point of the function and the (b) winding number profile indicate the existence of only periodic and quasi-periodic solutions. The winding number profile is computed in the line $\theta=0.45$. We set $q_0=3$, $q_1=6.0$ and $q_m=1.5$. }
\end{center}
\end{figure*}

Following the same procedure applied to the Tokamap in the last section, we construct the phase space and compute the winding number profile for two values of $K$. In FIG. \ref{rtm1}, we have the results for $K=0.5$.
	
For the results showed in FIG. \ref{rtm1}, we conclude that the revtokamap only presents regular solutions, for $K=0.5$. The phase space exhibit only islands and rotational curves and the winding number is defined for every $\psi \in [0,1]$. The winding number profile in  FIG. \ref{rtm1} (b) presents a nonmonotonic behavior, and a maximum value, indicated by the red symbol, around $\psi=0.5$. This point correspond to the shearless point mentioned before and it belongs to the shearless curve, also indicated in red in FIG. \ref{rtm1} (a).   The winding number plateaus highlighted by the orange and blue rectangles corresponds to the twin islands (isochronous solutions) of the same color in FIG. \ref{rtm1} (a). The twin islands present the same winding number, same period and each chain is located at one side of the shearless curve. We also observe the islands of period 3 (pink islands) with winding number $\omega_n=0.333...$. As identified in FIGS. \ref{wn_tm1} and \ref{wn_tm2}, the winding number of each island satisfy the relation $\omega_n=1/\tau$, where $\tau$ is the period of the island.
	
\begin{figure*}
\begin{center}
	\includegraphics[width=0.85\textwidth]{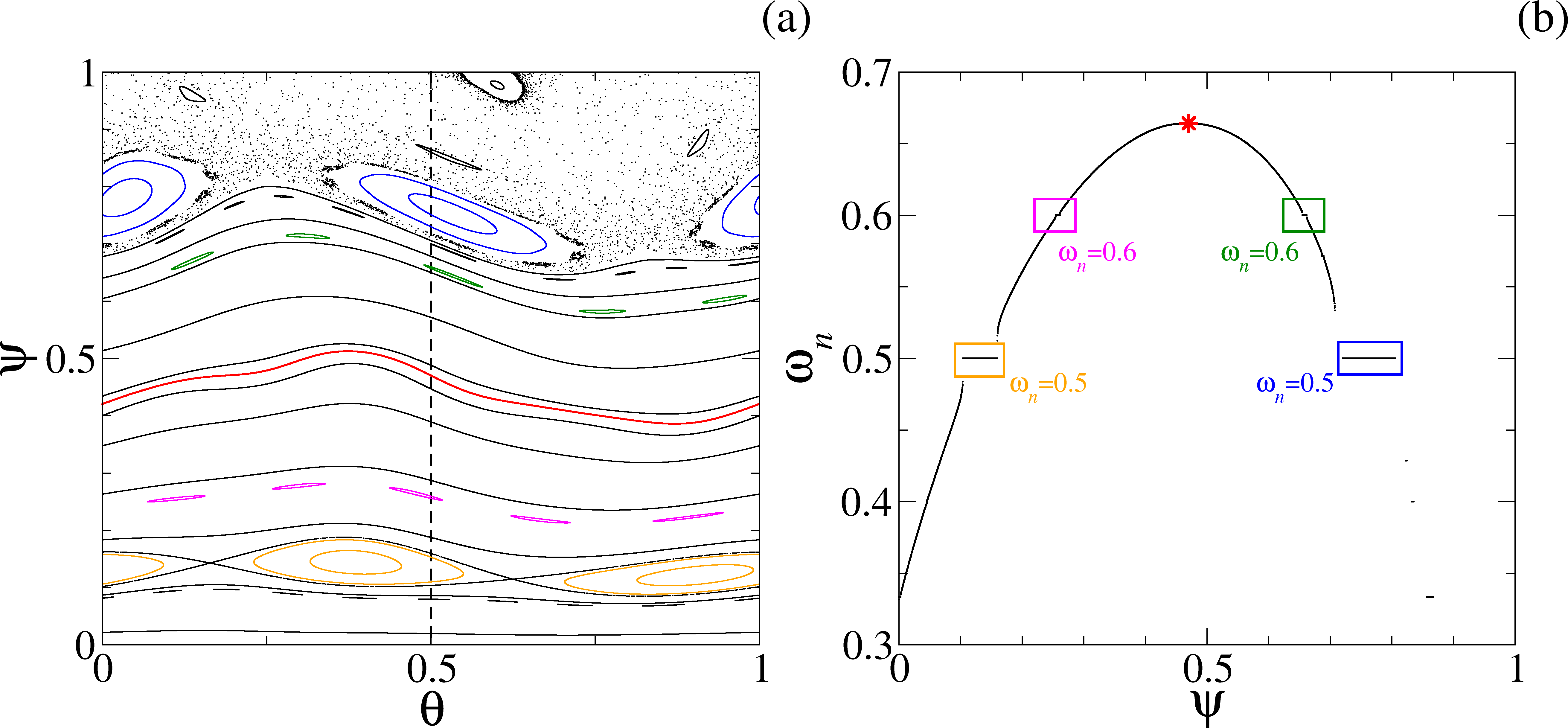}
	\caption{\label{rtm2} The Revtokmap for $K=2.0$, $q_0=3$, $q_1=6.0$ and $q_m=1.5$. In (a) we have the Poincaré section, and in (b) the winding number computed in $\theta=0.5$.}
\end{center}
\end{figure*}

Keeping the values for $q_0$, $q_1$ and $q_m$, in FIG. \ref{rtm2}, we have the Poincaré section and the winding number profile for $K=2.0$. From FIG. \ref{rtm2} (a), we observe that when the perturbation parameter $K$ is increased to $K=2.0$, some regular solutions at the upper region of the phase space are replaced by chaotic trajectories, indicated by the chaotic sea. The isochronous solutions of period 2, the blue and orange islands, remain and other two chain of islands are identified, the green and pink islands emerge. These last islands correspond to the plateaus in the winding number profile in FIG. \ref{rtm2} (b) highlighted by the same color. The isochronous solutions present $\omega_n=0.6=3/5$. The maximum value of $\omega_n$, highlighted by the red point in FIG. \ref{rtm2} (b), corresponds to the shearless curve, the red rational circle in FIG. \ref{rtm2} (a).

\section{Conclusions}
The Hamiltonian description of magnetic field lines is widely used for magnetic confined plasmas allowing the use of the powerful methods of Hamiltonian theory to interpret the results and characterize the dynamics regimes observed in experiments and computational simulations. The contributions of the Hamiltonian approach in plasmas physics range from the application of area-preserving maps, like the standard map, for the study of chaos \cite{chirikov}, to the Greene residue\cite{greene} and the Chirikov resonance overlap criterion\cite{escandePLA}, the nontwist systems, the renormalization group approach \cite{escandePR} and chaotic transport, just to name a few \cite{escande,escande2,evans}. Despite the importance and wide range of application, there are few elementary expositions on the subject. This paper attempts to fill this gap, presenting a tutorial of how the magnetic field lines are related to Hamiltonian systems with some representative application in toroidal plasmas.
	
Magnetic field lines are a non mechanical example of a system that can be described by the Hamiltonian formalism. From the variational principle we were able to present the description of field lines in confined plasmas for different coordinates and with the inclusion of a external perturbation. We also present applications of the description with the tokamap and revtokamap analysis. The examples presented here are simple, but they are paradigmatic for the study of confined plasmas and are adequate to demonstrate the Hamiltonian approach in a pedagogical form.
	
In order to emphasize the tutorial character of our paper, we provide the analytic calculations and the codes for the numerical simulations in the repository that can be found in \url{http://henon.if.usp.br/OscilControlData/HamiltonianDescriptionTutorial/}.

\section{Acknowledgements}
 We wish to acknowledge the support of the following Brazilian research agencies: Coordination for the Improvement of Higher Education Personnel (CAPES) under Grant No. 88887.320059/2019-00, 88881.143103/2017-01, the National Council for Scientific and Technological Development (CNPq - Grant No. 403120/2021-7, 301019/2019-3) and Fundação de Amparo à Pesquisa do Estado de São Paulo (FAPESP) under Grant No. 2022/12736-0, 2018/03211-6. %R. L. V. received partial financial support from the following Brazilian government agencies: CNPq (403120/2021-7, 301019/2019-3), CAPES (88881.143103/2017-01), and FAPESP (2022/04251–7).

\nocite{*}
%\bibliography{aipsamp}% Produces the bibliography via BibTeX.

\end{document}